\documentclass[journal,hidelinks]{IEEEtran}
\usepackage{scalerel}
\usepackage{amsmath}
\usepackage{booktabs,amssymb}
\usepackage{amsfonts}
\usepackage{soul}
\usepackage[most]{tcolorbox}
\usepackage{enumitem}
\usepackage{amsmath}
\usepackage{bm}
\usepackage{xcolor}
\usepackage{orcidlink}

\usepackage{graphicx} % For figure insertion
\usepackage{adjustbox}
\usepackage{wrapfig}
\usepackage{array}
\usepackage{multirow}
\usepackage{booktabs}
\usepackage{threeparttable}
\usepackage{xcolor} % if not already included

\usepackage{tablefootnote}
\usepackage{balance}
\usepackage{textcomp}
\usepackage{url}
\usepackage{xurl}
\usepackage{microtype}
\usepackage{pifont} 
\usepackage{algorithm}
\usepackage{array}
\usepackage[caption=false,font=normalsize,labelfont=sf,textfont=sf]{subfig}
\usepackage{textcomp}
\usepackage{stfloats}
\usepackage{url}

\usepackage{tikz}
\usetikzlibrary{positioning,calc,backgrounds}
\usepackage{tikzsymbols}
\usetikzlibrary{automata, positioning, shapes, fit, calc, shadows, shapes.geometric, arrows, arrows.meta, patterns, decorations.markings, decorations.pathreplacing, pgfplots.groupplots}
\usepackage{pgfplots}
\pgfplotsset{compat=1.18}
\usepackage{circuitikz}
\usepackage{tabularx,booktabs,array}
\usetikzlibrary{arrows.meta,positioning}

\usepackage{pgfplots}
\pgfplotsset{compat=1.18}
\usepgfplotslibrary{groupplots}

\usetikzlibrary{arrows.meta}
\usepackage{verbatim}
\usepackage{booktabs}
\usepackage{amsmath,amssymb}
\usepackage{algpseudocode} % from algorithmicx
\usetikzlibrary{arrows.meta,positioning,shapes,fit,calc}
\usepackage{graphicx}
\usepackage{tikz}
\usepackage{amsmath}
\usepgfplotslibrary{statistics}
\usepackage{filecontents}

\ifCLASSOPTIONcompsoc
  \usepackage[nocompress]{cite}
\else
  \usepackage{cite}
\fi
\ifCLASSINFOpdf
\else
\fi
\definecolor{orcidlogocol}{HTML}{A6CE39}
\tikzset{
    orcidlogo/.pic={
        \fill[orcidlogocol] svg{M256,128c0,70.7-57.3,128-128,128C57.3,256,0,198.7,0,128C0,57.3,57.3,0,128,0C198.7,0,256,57.3,256,128z};
        \fill[white] svg{M86.3,186.2H70.9V79.1h15.4v48.4V186.2z}
        svg{M108.9,79.1h41.6c39.6,0,57,28.3,57,53.6c0,27.5-21.5,53.6-56.8,53.6h-41.8V79.1z M124.3,172.4h24.5c34.9,0,42.9-26.5,42.9-39.7c0-21.5-13.7-39.7-43.7-39.7h-23.7V172.4z}
        svg{M88.7,56.8c0,5.5-4.5,10.1-10.1,10.1c-5.6,0-10.1-4.6-10.1-10.1c0-5.6,4.5-10.1,10.1-10.1C84.2,46.7,88.7,51.3,88.7,56.8z};
    }
}

\newcommand\orcidicon[1]{\href{https://orcid.org/#1}{\mbox{\scalerel*{
                \begin{tikzpicture}[yscale=-1,transform shape]
                \pic{orcidlogo};
                \end{tikzpicture}
            }{|}}}}

\def\BibTeX{{\rm B\kern-.05em{\sc i\kern-.025em b}\kern-.08em
    T\kern-.1667em\lower.7ex\hbox{E}\kern-.125emX}}

\begin{document}

\bstctlcite{IEEEexample:BSTcontrol}
\title{APQKE-DR: Entropy-Failure Analysis of a Post-Quantum-Initialized Ratcheted Channel}

\author{Muhammad~Asfand~Hafeez\textsuperscript{\orcidicon{0000-0001-8017-8447}}, \IEEEmembership{Graduate Student Member,~IEEE,}
        Zain~Ul~Abideen\textsuperscript{\orcidicon{0000-0002-8865-9402}}, \IEEEmembership{Member,~IEEE,},Amna~Zafar\textsuperscript{\orcidicon{0009-0009-4654-711X}}, Kazi~Hassan~Shakib\textsuperscript{\orcidicon{0000-0002-7894-0719}}, \IEEEmembership{Member,~IEEE}, Arslan~Munir\textsuperscript{\orcidicon{0000-0002-3126-8945}}, \IEEEmembership{Senior Member,~IEEE}
  %\thanks{Manuscript received 13 July 2022; revised 25 October 2022 and 3 February 2023; accepted 10 February 2023. This work has been partially conducted in the project ``ICT programme'' which was supported by the European Union through the ESF. This article was recommended by Associate Editor J. Rajendran. (Corresponding author: Zain Ul Abideen.)}
  \thanks{M.A. Hafeez, A. Zafar and A. Munir are with the Department of Computer Science and Electrical Engineering, Florida Atlantic University, 33431, Florida, USA (e-mail: mhafeez2024@fau.edu; azafar2025@fau.edu; arslanm@fau.edu).}
  \thanks{Zain Ul Abideen is with Department of Electrical and Computer Engineering, University of Idaho, Moscow, ID, USA (e-mail: \href{mailto:zabideen@uidaho.edu}{\mbox{\nolinkurl{zabideen@uidaho.edu}}}).}
  \thanks{K. H. Shakib is with Department of Computer Science, Kansas State University, Kansas, USA (e-mail: kshakib@ksu.edu).}
%\vspace*{-1.0cm}
}

% The paper headers
%\markboth{IEEE TRANSACTIONS ON INFORMATION FORENSICS AND SECURITY%, VOL.~43, NO.~12, DECEMBER~2022
%}{Muhammad~Asfand \MakeLowercase{\textit{et al.}}: APQKE-DR: Entropy-Failure Analysis of a Post-Quantum-Initialized Ratcheted Channel}
%\author{\IEEEauthorblockN{Anonymous Submission}}

% make the title area
\maketitle

% As a general rule, do not put math, special symbols or citations
% in the abstract

\begin{abstract}
Long-lived secure channels for embedded and cyber-physical systems (CPSs) must resist quantum attacks and failures in endpoint randomness. While several post-quantum key exchange protocols exist, their security assumes reliable randomness at the endpoint, but this assumption is not always true in embedded devices. %In this paper, we present a parameterized Entropy Attack Suite (EAS) for evaluating secure-channel implementations under two controlled endpoint-randomness failure modes: low-entropy initialization during early boot and random number generator (RNG) path subversion. As a case study, we design and implement APQKE-DR, a post-quantum-initialized secure channel. Its APQKE handshake authenticates two ephemeral ML-KEM~\cite{alagic2024module} exchanges using ML-DSA \cite{dang2024module} and binds the derived keys to the authenticated handshake transcript. The resulting initial root key ($RK_0$) then initializes an X25519 Double Ratchet for long-lived communication. We analyze the performance and security of APQKE-DR for smart grid communication, a particularly challenging application domain characterized by long-lived and resource-limited nodes, untrusted links, and frequent reboots. We also integrate APQKE-DR beneath the ANSI C12.22 and DNP3 smart grid protocols without modifying their application payload formats.
In this paper, we introduce the Entropy Attack Suite (EAS) to evaluate secure channels facing endpoint randomness failures, such as low-entropy boot and Random Number Generation (RNG) path subversion. As a case study, we design and implement APQKE-DR, a post-quantum secure channel. Its handshake uses ML-KEM~\cite{alagic2024module} and ML-DSA~\cite{dang2024module}, producing a root key for an X25519 Double Ratchet. We assess APQKE-DR’s performance and security for smart grid communication, and integrate it beneath ANSI C12.22 and DNP3 without altering payload formats.

On a Raspberry Pi Compute Module 5, the APQKE handshake exchanges 9,610 bytes with median latencies of 1.99 ms (node) and 0.82 ms (server). The ratcheted channel adds 31 bytes per message. With fresh randomness and ideal cryptographic primitives, ProVerif verifies the modeled secrecy and authentication properties. If randomness fails, EAS can recover the RNG stream and root key from public transcripts. Under RNG-path subversion, a weak generator exposes the root key across sessions. If subversion also affects X25519 keys, attackers may access ratchet keys and plaintexts. TPM seeding prevents early-boot attacks but not later subversion, while measured boot can block modified firmware. Secure deployment thus requires integrity along the entire entropy path, not just strong cryptographic primitives.

%On a Raspberry Pi Compute Module 5, the APQKE handshake exchanges 9,610 payload bytes and completes with median latencies of 1.990 ms (smart grid node) and 0.821 ms (server). The ratcheted channel adds 31 bytes per message in steady state. With healthy randomness, ProVerif in the Dolev–Yao model verifies APQKE-DR handshake secrecy, mutual authentication, transcript agreement, and message integrity. If endpoint randomness fails, EAS can recover the endpoint RNG stream and $RK_0$ from public handshake transcripts without breaking the primitives. Under RNG-path subversion, a diagnostic 16-bit linear congruential generator (LCG) yields only $2^8$ observable streams, enabling near-certain transcript-guided $RK_0$ recovery across sessions. If subversion also controls X25519 bootstrap or ratchet key generation, the attacker can expose ratchet chain keys and application plaintexts. Trusted Platform Module (TPM) seeding prevents early-boot recovery but not later RNG-path subversion, whereas measured boot tied to TPM platform configuration registers (PCRs) rejects modified firmware before channel setup. Overall, secure deployment requires integrity of the entire entropy path, not just post-quantum primitives.

\end{abstract}
% no keywords

% For peer review papers, you can put extra information on the cover
% page as needed:
% \ifCLASSOPTIONpeerreview
% \begin{center} \bfseries EDICS Category: 3-BBND \end{center}
% \fi
%
% For peerreview papers, this IEEEtran command inserts a page break and
% creates the second title. It will be ignored for other modes.
\IEEEpeerreviewmaketitle

%===========================================
% Introduction
%===========================================
\vspace{-5mm}
\section{Introduction}
\vspace{-2mm}

Modern embedded and cyber-physical systems (CPSs) rely on continuous communication between field devices and centralized control services, carrying measurements to control software and returning commands that drive system operation. Smart grids apply this model to electricity distribution: smart meters and other field devices send usage and grid measurements to supervisory control and data acquisition (SCADA) backends, which monitor conditions and coordinate control actions \cite{gallegos2024sustainable, amin2008challenges, kabalci2018introduction}. This connectivity improves visibility and automation but exposes devices and communication channels to attacks that can cause incorrect billing, falsified measurements, loss of visibility, or disruption of control functions \cite{gunduz2020cyber, goel2015security, inl2016cyber}.

In these systems, each connection has two \emph{endpoints}: a field device and a backend server. These endpoints establish session keys to protect their communication, typically using Rivest-Shamir-Adleman (RSA) or elliptic-curve cryptography (ECC) \cite{yan2022overview}. A sufficiently capable quantum adversary can break these public-key primitives and recover these keys, which would allow it to decrypt previously recorded traffic \cite{wong2023shor, giron2023migrating}. Migrating away from these algorithms is difficult because field devices span a wide capability spectrum, from microcontroller unit (MCU)-class platforms to Linux-based gateways and the most resource-constrained devices in this range offer limited memory capacity and computational headroom.

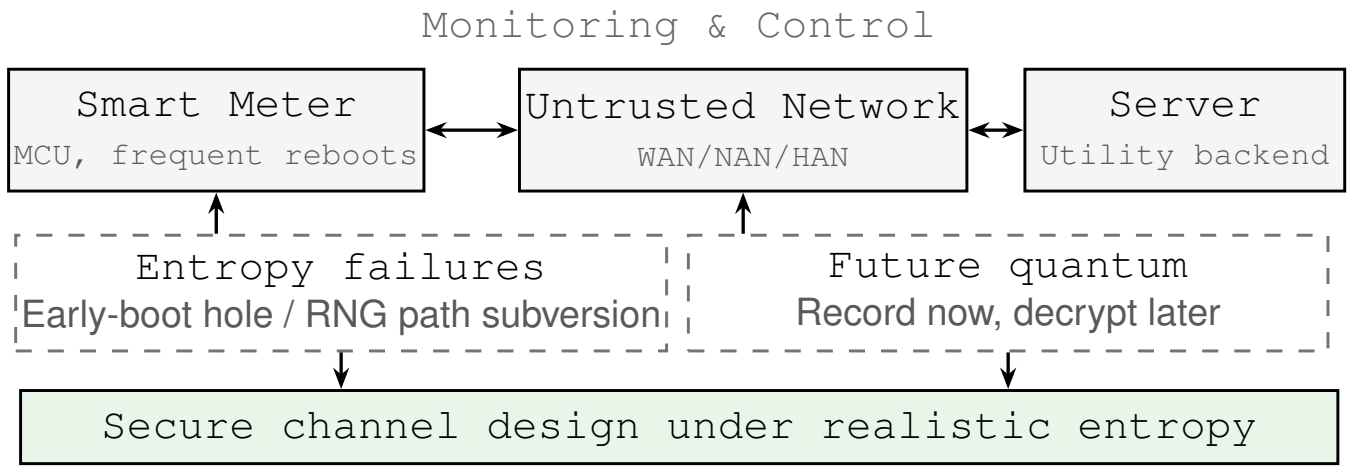
\begin{figure}[htb]
\centering
\begingroup
\definecolor{introGray}{RGB}{245,245,245}
\definecolor{introGreen}{RGB}{232,245,233}
\newcommand{\introDetail}[1]{%
  {\sffamily\fontsize{7}{8}\selectfont\textcolor{black!65}{#1}}}
% Shared size for the subtitles in the three top boxes.
\newcommand{\introTopDetail}[1]{%
  {\ttfamily\fontsize{6}{7}\selectfont\textcolor{black!65}{#1}}}
% Horizontal positions in mm. The network and its arrows move together.
\def\introMeterX{13.1}
\def\introNetworkX{47.7}
\def\introServerX{76.7}
\def\introEntropyX{21.3}
\def\introQuantumX{65.1}

\resizebox{\columnwidth}{!}{%
\begin{tikzpicture}[
  x=1mm,y=1mm,>={Stealth[length=1.4mm,width=1.05mm]},
  font=\ttfamily\fontsize{8}{9}\selectfont,
  box/.style={draw=black,fill=introGray,align=center,
    line width=.55pt,inner sep=1pt,minimum height=8mm},
  threat/.style={box,draw=black!55,dashed,fill=white,
    minimum height=7.6mm},
  flow/.style={line width=.62pt}]

% Context row.
\node[font=\ttfamily\fontsize{8}{9}\selectfont,text=black!60,
      inner sep=0pt] at (43.5,27) {Monitoring \& Control};

\node[box,minimum width=26mm] (C) at (\introMeterX,20.2)
  {Smart Meter\\\introTopDetail{MCU, frequent reboots}};
\node[box,minimum width=27mm] (N) at (\introNetworkX,20.2)
  {Untrusted Network\\\introTopDetail{WAN/NAN/HAN}};
\node[box,minimum width=21mm] (S) at (\introServerX,20.2)
  {Server\\\introTopDetail{Utility backend}};

\draw[flow,<->] (C.east) -- (N.west);
\draw[flow,<->] (N.east) -- (S.west);

% Wider threat boxes keep each description on one line.
\node[threat,minimum width=41.2mm] (E) at (\introEntropyX,9.5)
  {Entropy failures\\\introDetail{Early-boot hole / RNG path subversion}};
\node[threat,minimum width=42mm] (Q) at (\introQuantumX,9.5)
  {Future quantum\\\introDetail{Record now, decrypt later}};

% Use vertical connectors even when the box centers differ.
\draw[flow,->] (C.south |- E.north) -- (C.south);
\draw[flow,->] (N.south |- Q.north) -- (N.south);

\node[box,fill=introGreen,minimum width=86.5mm,
      minimum height=4.8mm] (W) at (43.5,0.7)
  {Secure channel design under realistic entropy};
\draw[flow,->] (E.south) -- (E.south |- W.north);
\draw[flow,->] (Q.south) -- (Q.south |- W.north);

\end{tikzpicture}%
}
\endgroup
\vspace{-6mm}
\caption{Smart meters communicate with SCADA over untrusted networks,
facing quantum record-now-decrypt-later risk and practical entropy failures.}
\label{fig:intro_motivation}
\vspace{-6mm}
\end{figure}

Post-quantum key exchange (PQKE) addresses the quantum threat, but it still requires unpredictable random inputs at the endpoints \cite{piscopo2024design}. For a key-encapsulation mechanism (KEM), randomness is needed to generate an ephemeral keypair and to establish a shared secret using the recipient's public key. An endpoint's random number generator (RNG) can violate this requirement if it is initialized with too few unpredictable bits or if modified software replaces its output with a predictable sequence \cite{han12026min}. Suppose the device uses the same RNG stream for its random KEM inputs and its handshake nonce which is used to distinguish protocol runs. If the set of possible streams is small enough to enumerate, an attacker first regenerates the nonce for each stream and compares it with the recorded nonce. For each match, the attacker regenerates the device's KEM keypair and encapsulation ciphertext. The regenerated public key and ciphertext are then compared with the recorded handshake messages. If both match, the attacker uses the regenerated private key and encapsulation randomness to recover the shared secrets. This search requires knowledge of the RNG algorithm and the order in which the protocol requests random values, but no further interaction with the device \cite{rfc8446}. Fig.~\ref{fig:intro_motivation} illustrates the \textbf{quantum threat and endpoint randomness failures} in communication between a smart meter and a utility backend.

Existing protocols provide an authenticated key establishment mechanism for updating encryption keys. Transport Layer Security (TLS)~1.3 authenticates its handshake transcript and confirms the derived keys, but its \texttt{KeyUpdate} mechanism derives each new traffic secret from the previous one. An attacker who obtains that secret can therefore derive subsequent traffic keys on the same connection \cite{rescorla2018transport}. The Signal protocol \cite{cohn2020formal} addresses this limitation by incorporating fresh shared secrets through ratcheted messaging, including classical and post-quantum updates in its Triple Ratchet \cite{perrin2016double}. However, its post quantum (PQ) extended Diffie-Hellman (PQXDH) handshake still relies on classical elliptic-curve cryptography for mutual authentication \cite{kret2023pqxdh}.

% ---------------------- INTRO FIGURE (motivational overview) ----------------------
% If these styles are already defined elsewhere in your paper, you can remove them here.
\tikzstyle{title} = [fill=white,font=\fontsize{8}{8}\color{black!50}\ttfamily]
\tikzstyle{block} = [draw, fill=white, rectangle, align=center, minimum height=1em, minimum width=1em, font=\fontsize{8}{7}\color{black}\ttfamily]
\tikzstyle{smallblock} = [draw, fill=white, rectangle, align=center, minimum height=1em, minimum width=1em, font=\fontsize{8}{7}\color{black}\ttfamily]
\tikzstyle{specialblock} = [draw, fill=white, rectangle, align=center, minimum height=1em, minimum width=1em, font=\fontsize{8}{7}\color{black}\ttfamily]
\definecolor{secure}{RGB}{232,245,233} % light green
\definecolor{Grey}{RGB}{245,245,245} % light gray

%\introrev{To address these authentication and recovery requirements, we design APQKE-DR, which combines an authenticated post-quantum key exchange with a Double Ratchet (DR). The channel uses pre-provisioned authentication keys for direct node-to-server communication, and the DR incorporates fresh X25519 shared secrets into its key updates.} Using a Raspberry Pi Compute Module 5 as a gateway-class embedded Linux endpoint, we \introrev{evaluate the channel with the} \emph{Entropy Attack Suite (EAS)} under two controlled \introrev{randomness} failures. In the \emph{early-boot entropy-hole} scenario, \introrev{we limit the unpredictable bits used to seed the RNG during boot.} In the \emph{RNG-path subversion} scenario, modified firmware \introrev{replaces RNG outputs with a deterministic stream generated from a small internal state.} \introrev{The experiments target recovery of the initial handshake root key $RK_0$ from recorded transcripts. Recovering $RK_0$ alone does not reveal message keys to a passive classical attacker when DR initialization also incorporates an X25519 shared secret unknown to that attacker. We separately analyze the additional message-key exposure when RNG subversion extends to X25519 key generation.} APQKE-DR is post-quantum initialized, but its X25519 ratchet offers only classical post-compromise recovery. \introrev{After the attacker loses access to the channel state, later messages can become confidential again if a key update introduces a shared secret unknown to the attacker.} Our contributions are as follows:

To address these authentication and recovery requirements, we design APQKE-DR, which combines an authenticated post-quantum key exchange with a Double Ratchet (DR) for direct node-to-server communication. Using a Raspberry Pi Compute Module 5 as a gateway-class embedded Linux endpoint, we evaluated the channel and hardware-rooted defenses with the \emph{Entropy Attack Suite (EAS)} under two controlled randomness failures: early-boot entropy holes and RNG-path subversion. APQKE-DR is post-quantum initialized, but its X25519 ratchet offers only classical post-compromise recovery. After the attacker loses access to the channel state, later messages can become confidential again if a key update introduces a shared secret unknown to the attacker. Our contributions are as follows:

%\vspace{-1mm}
\begin{itemize}
      \item We design and implement \textbf{APQKE-DR}, an experimental secure channel for the smart grid node-to-server interface. The APQKE handshake provides authenticated post-quantum key establishment using ML-KEM for key encapsulation and ML-DSA for signatures \cite{nistfips203,nistfips204}, binds the derived root key to an authenticated transcript hash, and applies explicit \textit{Finished-style} key confirmation \cite{rfc8446,dowling2015cryptographic}. The initial root key $RK_0$ initializes the DR data channel, cryptographically linked to the authenticated handshake transcript for subsequent messaging.

      \item We introduce an \emph{EAS} to model early-boot entropy holes and RNG-path subversion, demonstrating offline recovery of the root key ($RK_0$) from public transcripts and enabling evaluation of the handshake-plus-channel protocol under degraded entropy conditions rather than idealized randomness assumptions \cite{nistsp80090b}.
    
      \item We compare Trusted Platform Module (TPM) seeding with Platform Configuration Register (PCR)-bound measured-boot gating. TPM seeding blocks recovery in the tested early-boot configurations but does not stop replacement of RNG outputs after seeding. Under the enforced boot policy, PCR-bound gating blocks session establishment for the tested modified firmware.
    
      \item Experimental evaluation measures latency, bandwidth, and total system power on the gateway-class testbed, and integrates APQKE-DR within ANSI C12.22~\cite{shees2025performance} and DNP3~\cite{clarke2004practical} without changing their application payload formats. We model handshake secrecy, authentication, and channel binding in ProVerif~\cite{blanchet2018proverif} against an active network attacker, using ideal cryptographic primitives and fresh randomness.
\end{itemize}
%\vspace{-1mm}
The paper is organized as follows: Section~\ref{sec:background} reviews background and related work. Section~\ref{sec:methodology} presents the APQKE-DR protocol. Section~\ref{sec:EAS} introduces EAS and its implementation. Section~\ref{sec:sec-eval} covers the security analysis and results. Finally, Section~\ref{sec:con} concludes the paper.

% \cite{alwen2019double}
%===========================================
% Background and System Context
%===========================================
\vspace{-2mm}
\section{Background, System and Threat Model}\label{sec:background}
\vspace{-2mm}
The smart grid connects resource-constrained field devices via heterogeneous links. Key establishment and authentication protocols in smart grids focus on minimizing computational, bandwidth, and energy costs at endpoints, often using elliptic-curve primitives such as ECDH/ECDSA, along with symmetric encryption and message authentication codes (MACs)~\cite{nkenyereye2025lightweight}. A wide range of lightweight and Advanced Metering Infrastructure (AMI) focused designs refine handshake structure, reduce state, and optimize message formats to fit metering and SCADA constraints \cite{zafar2025lseg, cheng2024llra}.

Motivated by ``record now, decrypt later'' threats and the susceptibility of RSA/ECC to quantum algorithms \cite{wong2023shor, giron2023migrating}, recent work has explored integrating PQ primitives into smart grid and AMI security architectures. Prototypes have investigated Kyber and FrodoKEM-like mechanisms and their feasibility for embedded AMI deployments \cite{lagrota2025quantum, da2021soc}, while other proposals develop lattice-based mutual authentication and key exchange protocols designed for smart grid communication \cite{shekhawat2024quantum}. These studies demonstrate that PQ primitives can be engineered into metering infrastructures, but they predominantly focus on initial session establishment and system integration. End-to-end secure messaging over long-lived channels, including forward secrecy, post-compromise recovery, and robustness to realistic endpoint randomness failures, is less frequently discussed.

Outside the smart grid literature, secure-channel protocols provide well-established patterns for transcript binding and explicit key confirmation, notably in TLS~1.3 (transcript hash and Finished) \cite{rfc8446, dowling2015cryptographic}, as well as modular handshake frameworks, such as noise \cite{noise2018framework}. For long-lived encrypted messaging, the Signal Double Ratchet is a widely deployed construction for forward secrecy and post-compromise recovery, with formal security analyses \cite{alwen2019double, cremers2023sessionhandling}. Parallel work has also explored post-quantum directions for secure-messaging handshakes \cite{brendel2022pqxdh}.

%===========================================
% System and Threat Model
%===========================================
\vspace{-2mm}
\subsection{System and Threat Model}\label{sec:threat}
\vspace{-2mm}

%We consider a smart grid where field nodes $\{SGN_i\}$ establish encrypted APQKE-DR channels with a backend server ($S$) in a controlled utility environment (Fig.~\ref{fig:system_model}). Smart meters report electricity use, remote terminal units (RTUs) collect measurements and relay equipment commands, phasor measurement units (PMUs) provide time-synchronized measurements, and gateways collect local data. AMI concentrators forward encrypted traffic between multiple nodes and $S$ without access to the channel keys. Field links may include home area networks (HANs) and neighborhood area networks (NANs); backhaul links connect concentrators to $S$. The links are untrusted: an attacker may observe, modify, or block the transmitted messages.

Smart-grid nodes $\{SGN_i\}$ use encrypted APQKE-DR channels to communicate with the utility server $S$ (Fig.~\ref{fig:system_model}). Field devices include smart meters, remote terminal units (RTUs), phasor measurement units (PMUs), and gateways. AMI concentrators relay ciphertext without access to channel keys. Field links may use home or neighborhood area networks, such as home area networks (HANs) and neighborhood area networks (NANs). Both field and backhaul links are untrusted: an attacker can observe, modify, or block messages.

A trusted authority provisions each node with the server's ML-DSA verification key and registers each node's verification key at $S$. Handshakes do not contact the authority. Nodes may have a hardware entropy source or TPM. Software updates are signed, and legitimate ratchet-state loss after a reset requires a fresh APQKE handshake. We distinguish six adversaries:

\vspace{-1mm}
\begin{itemize}
  \setlength{\itemsep}{1pt}
  \setlength{\topsep}{2pt}

  \item $\mathcal{A}_{\mathrm{net}}$ is an active Dolev-Yao network adversary that can observe, modify, inject, replay, reorder, delay, or drop messages; initiate concurrent sessions; and mount computational or state-exhaustion denial-of-service attempts.

  \item $\mathcal{A}_{\mathrm{seed}}$ can reset or power-cycle $SGN_i$, control handshake timing, and knows the modeled low-entropy boot-seed distribution but cannot read endpoint secrets.

  \item $\mathcal{A}_{\mathrm{rng}}$ controls only the bytes returned by the node's RNG application programming interface (API). It cannot access long-term keys, plaintext, ratchet state, TPM state, or other process memory.

  \item $\mathcal{A}_{\mathrm{img}}$ can install and attempt to boot a modified image but cannot forge an approved measurement or bypass the enforced TPM PCR policy. An unauthorized image, therefore, causes the node to fail closed before channel establishment.

  \item $\mathcal{A}_{\mathrm{state}}$ temporarily obtains the current Double Ratchet state, including the current root and chain keys, but later loses access and does not obtain provisioned keys or future randomness.

  \item $\mathcal{A}_{\mathrm{Q}}$ records network traffic today and gains access to a large-scale quantum computer in the future, without live endpoint compromise.
\end{itemize}
\vspace{-1mm}

We assume a trustworthy provisioning authority and ML-DSA keys, correct cryptographic implementations, secure erasure of ephemeral KEM secrets, and used message keys, enforcement of the TPM measured-boot policy, and integrity of persistent credentials and ratchet state. Legitimate state loss is allowed, but malicious rollback or cloning of persistent state is out of scope, as are arbitrary or ongoing runtime-memory compromise, invasive physical attacks, compromise of the trusted authority or TPM, and side-channel or fault attacks. The protocol does not aim to resist sustained flooding, total state exhaustion, or indefinite message dropping.

$\mathcal{A}_{\mathrm{Q}}$ and $\mathcal{A}_{\mathrm{state}}$ capture different goals. $\mathcal{A}_{\mathrm{Q}}$ addresses confidentiality of recorded sessions initialized through ML-KEM against future passive quantum adversaries. $\mathcal{A}_{\mathrm{state}}$ addresses recovery after live ratchet-state compromise, which depends on a later uncompromised X25519 update and therefore offers only classical post-compromise recovery.

% ---------------------- FIGURE: System model (detailed, vertical, clean) ----------------------
% Uses your existing styles: title / block / specialblock / smallblock.

\definecolor{cSGN}{RGB}{255,244,230} % light orange

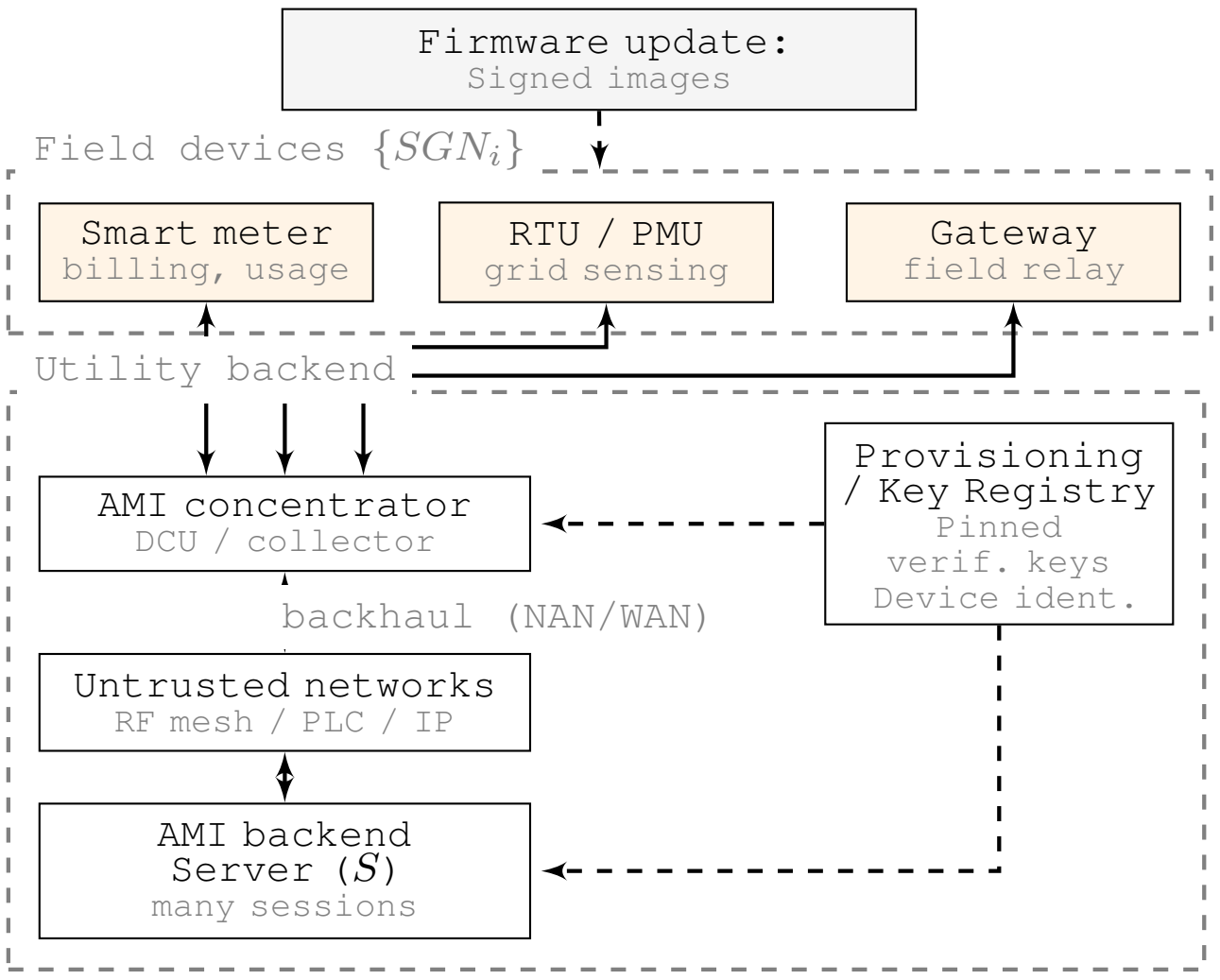
\begin{figure}[htb]
\centering
\begin{tikzpicture}[auto, >=latex']

% -------------------- Field device types (top row) --------------------
\node[specialblock, fill= cSGN, text width=2.1cm] (m) at (-2.55, 0.10)
{Smart meter\\
{\fontsize{7}{8}\selectfont\textcolor{gray}{billing, usage}}};

\node[specialblock, fill= cSGN, text width=2.1cm] (r) at (0.25, 0.10)
{RTU / PMU\\
{\fontsize{7}{8}\selectfont\textcolor{gray}{grid sensing}}};

\node[specialblock, fill= cSGN, text width=2.1cm] (g) at (3.10, 0.10)
{Gateway\\
{\fontsize{7}{8}\selectfont\textcolor{gray}{field relay}}};

% Optional dashed grouping for the fleet
\node[draw, dashed, thick, color=gray, inner xsep=6pt, inner ysep=6pt,
      fit=(m)(r)(g)] (field) {};
\node[title, anchor=west] at ([xshift=2pt,yshift=4pt]field.north west)
{Field devices $\{SGN_i\}$};

% -------------------- Firmware update channel (left, points to field devices) --------------------
\node[block, fill=Grey, text width=4.2cm] (fw) at (0.20, 1.45)
{Firmware update:
{\fontsize{7}{8}\selectfont\textcolor{gray}{Signed images}}};

\draw[thick, dashed, ->, shorten <=0pt, shorten >=0pt]
(fw.south) -- ([xshift=-2pt]field.north);

% -------------------- AMI concentrator (below field) --------------------
\node[block, text width=3.2cm] (dcu) at (-2.00, -1.8)
{AMI concentrator\\
{\fontsize{7}{8}\selectfont\textcolor{gray}{DCU / collector}}};

% Links from device types to concentrator (land on different points to avoid overlap)
\draw[thick,<->,shorten <=0pt,shorten >=0pt,rounded corners=0pt]
  (r.south)  |- ([yshift=9mm]dcu.north) -- (dcu.north);

\draw[thick,<->,shorten <=0pt,shorten >=0pt,rounded corners=0pt]
  (m.south)  |- ([yshift=6mm,xshift=-0.55cm]dcu.north) -- ([xshift=-0.55cm]dcu.north);

\draw[thick,<->,shorten <=0pt,shorten >=0pt,rounded corners=0pt]
  (g.south)  |- ([yshift=7mm, xshift=0.55cm]dcu.north) -- ([xshift=0.55cm]dcu.north);

%\node[title, align=center] at (0.50, -0.65) {Field links (HAN/NAN)};

% -------------------- Untrusted networks (below concentrator) --------------------
\node[block, text width=3.2cm] (net) at (-2.0, -3.05)
{Untrusted networks\\
{\fontsize{7}{8}\selectfont\textcolor{gray}{RF mesh / PLC / IP}}};

\draw[thick, <->, shorten <=0pt, shorten >=0pt] (dcu.south) -- (net.north);
\node[title, align=center] at (-0.50, -2.450) {backhaul (NAN/WAN)};

% -------------------- Backend server (bottom) --------------------
\node[specialblock, text width=3.2cm] (srv) at (-2.0, -4.23)
{AMI backend Server ($S$)\\
{\fontsize{7}{8}\selectfont\textcolor{gray}{many sessions}}};

\draw[thick, <->, shorten <=0pt, shorten >=0pt] (net.south) -- (srv.north);

% -------------------- Provisioning / PKI (right side) --------------------
\node[block, text width=2.2cm] (pki) at (3.00, -1.8)
{Provisioning / Key Registry\\
{\fontsize{7}{8}\selectfont\textcolor{gray}{Pinned verif. keys Device ident.}}};

% Dashed arrows (routed cleanly, no crossings through boxes)
\draw[thick, dashed, ->, shorten <=0pt, shorten >=0pt]
(pki.west) -- ([xshift=2pt]dcu.east);

\draw[thick, dashed, ->, shorten <=0pt, shorten >=0pt]
(pki.south) -- ++(0,-0.25) |- ([xshift=2pt]srv.east);

% Optional dashed grouping for the backend side
\node[draw, dashed, thick, color=gray, inner xsep=6pt, inner ysep=6pt,
      fit=(srv)(pki)] (util) {};
\node[title, anchor=west] at ([xshift=2pt,yshift=4pt]util.north west)
{Utility backend};

\end{tikzpicture}
\vspace{-2mm}
\caption{$\{SGN_i\}$ connect to backend Server through field links and an AMI concentrator. Provisioning distributes trust anchors and registered verification keys, while signed updates modify device software.}
\label{fig:system_model}
\vspace{-5mm}
\end{figure}

%===========================================
% Proposed Protocols
%===========================================
\vspace{-2mm}
\section{Proposed Protocols} \label{sec:methodology}
\vspace{-2mm}

In this section, we present our APQKE-DR protocol for secure communication between a provisioned embedded endpoint $SGN_i$ and the utility backend server $S$. The design includes two parts: \textbf{APQKE}, which is an authenticated post-quantum handshake generating an initial root key ($RK_0$) and an authenticated transcript hash ($TH_1$), and the \textbf{DR} data channel, which is resource-aware, seeded by $RK_0$, and tied to the handshake using $TH_1$ for application traffic.
%To manage resources, we have established a resource-aware rekey schedule. APQKE is executed during the initial connection, in case of state loss, and periodically thereafter. Meanwhile, DR operates continuously, incorporating one Diffie-Hellman (DH) ratchet refresh per session with compact message headers.

To manage resources,We have implemented a resource-aware rekey schedule. APQKE is executed during the initial connection, during state loss, and periodically thereafter. Meanwhile, the Diffie-Hellman (DH) ratchet refresh operates continuously with compact message headers.

%------------------------------------------------------------
\vspace{-3mm}
\subsection{Trusted Authority (TA) assisted distribution of long-term ML-DSA keys}
\label{sec:pki}
\vspace{-2mm}
We assume a trusted authority (TA) that distributes long-term ML-DSA public keys out-of-band before deployment. Unlike certificate-based PKI, the TA does not issue certificates or transmit certificate material during the handshake. Instead, the TA securely provisions the server's long-term signature verification key to each device and vice versa.

In concrete terms, the backend server has a long-term ML-DSA key pair $(sk^{sig}_S, pk^{sig}_S)$, and each node of the smart grid has $(sk^{sig}_{SGN_i}, pk^{sig}_{SGN_i})$. During manufacturing, $pk^{sig}_S$ is written in each $SGN_i$ (key pin), while the server stores $(ID_i, pk^{sig}_{SGN_i})$ in a registration table indexed by device identity. This enables APQKE to authenticate all ephemeral ML-KEM keys by verifying signatures under pre-provisioned ML-DSA verification keys, without repeatedly transmitting any static public keys on the network.

For compact on-wire identity representation, deployments may use a short key identifier, e.g.,\(kid_S = H(pk^{sig}_S), kid_i = H(pk^{sig}_{SGN_i})\), so that a receiver can select the correct pinned/registered key with minimal overhead.

%============================================================
%APKQE
%============================================================

\vspace{-2mm}
\subsection{APQKE Handshake}
\label{sec:proto1}
\vspace{-2mm}

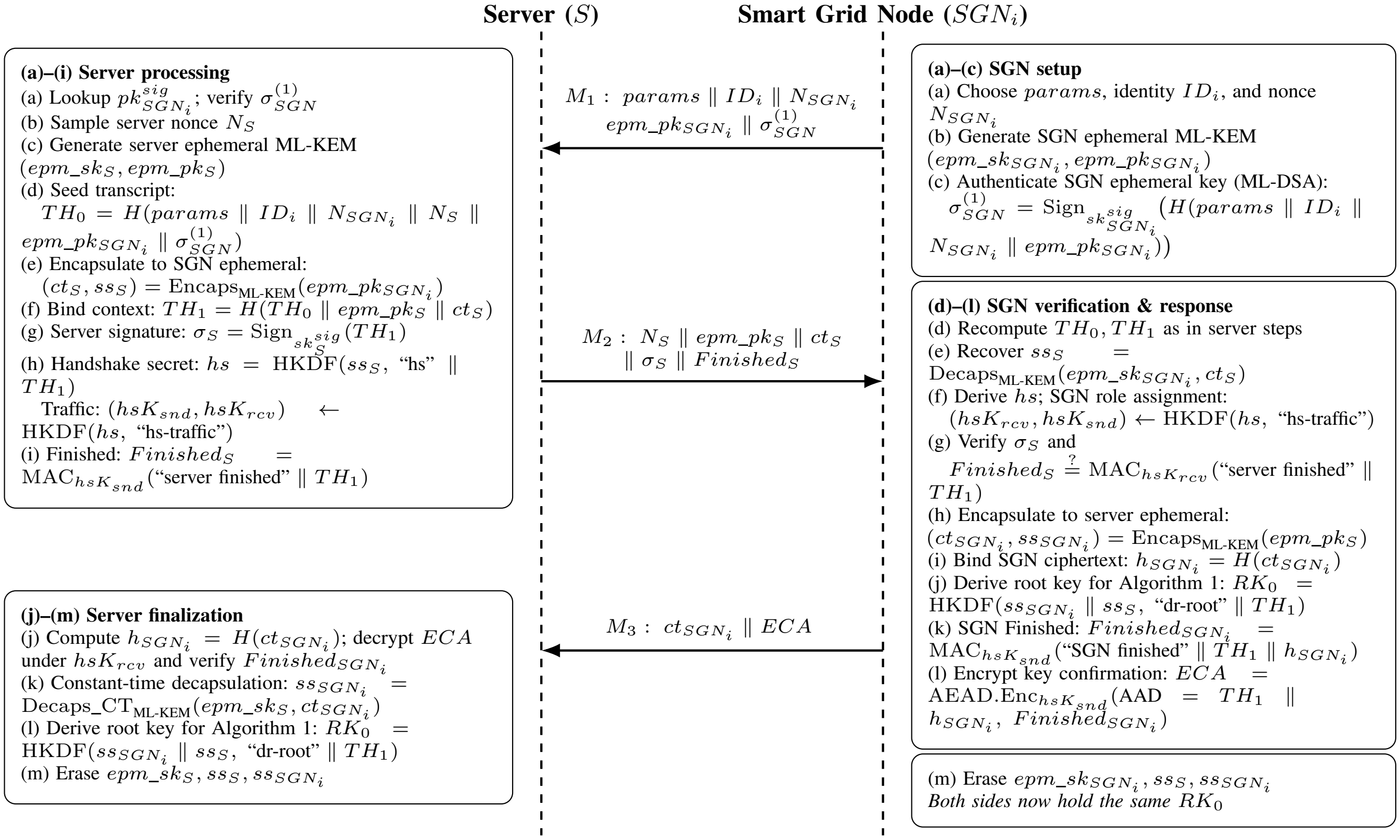
\begin{figure*}[t]
\centering
\begin{tikzpicture}[
font=\scriptsize,
>=Latex,
lifeline/.style={dashed, thick},
box/.style={draw, rounded corners, align=left, inner sep=6pt},
msgarrow/.style={->, thick}
]

% ---- geometry (tune these only if needed) ----
\def\xS{0}         % Server x
\def\xC{4.25}       % SGN x
\def\TopY{0.4}
\def\BotY{-9.8}

% ---- lifeline headers ----
\node[font=\small] (Shead) at (\xS,\TopY)
{\textbf{Server ($S$)}};

\node[font=\small] (Chead) at (\xC,\TopY)
{\textbf{Smart Grid Node ($SGN_i$)}};

% ---- dashed lifelines ----
\draw[lifeline] (\xS,0.2) -- (\xS,\BotY);
\draw[lifeline] (\xC,0.2) -- (\xC,\BotY);

% =======================
% Boxes
% =======================

% (a)-(c) SGN setup (TOP RIGHT)
\node[box, text width=5.6cm, anchor=north west]
(C1) at (\xC+0.35,0.05) {%
\textbf{(a)--(c) SGN setup}\\
(a) Choose $params$, identity $ID_i$, and nonce $N_{SGN_i}$\\
(b) Generate SGN ephemeral ML-KEM
$(epm\_sk_{SGN_i}, epm\_pk_{SGN_i})$\\
(c) Authenticate SGN ephemeral key (ML-DSA):\\
\quad
$\sigma^{(1)}_{SGN}
=\mathrm{Sign}_{sk^{sig}_{SGN_i}}\!
\bigl(
H(params\parallel ID_i\parallel N_{SGN_i}
\parallel epm\_pk_{SGN_i})
\bigr)$
};

% (a)-(i) server processing (LEFT, below)
\node[box, text width=5.9cm, anchor=north east]
(S1) at (\xS-0.35,-0.00) {%
\textbf{(a)--(i) Server processing}\\
(a) Lookup $pk^{sig}_{SGN_i}$; verify
$\sigma^{(1)}_{SGN}$\\
(b) Sample server nonce $N_S$\\
(c) Generate server ephemeral ML-KEM
$(epm\_sk_S, epm\_pk_S)$\\
(d) Seed transcript:\\
\quad
$TH_0=H(params\parallel ID_i\parallel N_{SGN_i}
\parallel N_S\parallel epm\_pk_{SGN_i}
\parallel \sigma^{(1)}_{SGN})$\\
(e) Encapsulate to SGN ephemeral:\\
\quad
$(ct_S,ss_S)
=\mathrm{Encaps}_{\text{ML-KEM}}(epm\_pk_{SGN_i})$\\
(f) Bind context:
$TH_1=H(TH_0\parallel epm\_pk_S\parallel ct_S)$\\
(g) Server signature:
$\sigma_S=\mathrm{Sign}_{sk^{sig}_S}(TH_1)$\\
(h) Handshake secret:
$hs=\mathrm{HKDF}(ss_S,\ \text{``hs''}\parallel TH_1)$\\
\quad Traffic:
$(hsK_{snd},hsK_{rcv})
\leftarrow
\mathrm{HKDF}(hs,\ \text{``hs-traffic''})$\\
(i) Finished:
$Finished_S
=\mathrm{MAC}_{hsK_{snd}}
(\text{``server finished''}\parallel TH_1)$
};

% (d)-(l) SGN verification & response (RIGHT, below)
\node[box, text width=5.6cm, anchor=north west]
(C2) at (\xC+0.35,-2.9) {%
\textbf{(d)--(l) SGN verification \& response}\\
(d) Recompute $TH_0,TH_1$ as in server steps\\
(e) Recover
$ss_S
=\mathrm{Decaps}_{\text{ML-KEM}}
(epm\_sk_{SGN_i},ct_S)$\\
(f) Derive $hs$; SGN role assignment:\\
\quad
$(hsK_{rcv},hsK_{snd})
\leftarrow
\mathrm{HKDF}(hs,\ \text{``hs-traffic''})$\\
(g) Verify $\sigma_S$ and\\
\quad
$Finished_S
\stackrel{?}{=}
\mathrm{MAC}_{hsK_{rcv}}
(\text{``server finished''}\parallel TH_1)$\\
(h) Encapsulate to server ephemeral:
$(ct_{SGN_i},ss_{SGN_i})
=\mathrm{Encaps}_{\text{ML-KEM}}(epm\_pk_S)$\\
(i) Bind SGN ciphertext:
$h_{SGN_i}=H(ct_{SGN_i})$\\
(j) Derive root key for Algorithm~1:
$RK_0
=\mathrm{HKDF}
(ss_{SGN_i}\parallel ss_S,\
\text{``dr-root''}\parallel TH_1)$\\
(k) SGN Finished:
$Finished_{SGN_i}
=\mathrm{MAC}_{hsK_{snd}}
(\text{``SGN finished''}\parallel TH_1
\parallel h_{SGN_i})$\\
(l) Encrypt key confirmation:
$ECA
=\mathrm{AEAD.Enc}_{hsK_{snd}}
(\text{AAD}=TH_1\parallel h_{SGN_i},\
Finished_{SGN_i})$
};

% (j)-(m) server finalization (LEFT, below)
\node[box, text width=5.9cm, anchor=north east]
(S2) at (\xS-0.35,-6.75) {%
\textbf{(j)--(m) Server finalization}\\
(j) Compute $h_{SGN_i}=H(ct_{SGN_i})$;
decrypt $ECA$ under $hsK_{rcv}$ and verify
$Finished_{SGN_i}$\\
(k) Constant-time decapsulation:
$ss_{SGN_i}
=\mathrm{Decaps\_CT}_{\text{ML-KEM}}
(epm\_sk_S,ct_{SGN_i})$\\
(l) Derive root key for Algorithm~1:
$RK_0
=\mathrm{HKDF}
(ss_{SGN_i}\parallel ss_S,\
\text{``dr-root''}\parallel TH_1)$\\
(m) Erase $epm\_sk_S,ss_S,ss_{SGN_i}$
};

% SGN erase (BOTTOM RIGHT, small)
\node[box, text width=5.6cm, anchor=north west]
(C3) at (\xC+0.35,-8.78) {%
(m) Erase $epm\_sk_{SGN_i},ss_S,ss_{SGN_i}$\\
\textit{Both sides now hold the same $RK_0$}
};

% =======================
% Message arrows
% =======================

\draw[msgarrow] (\xC,-1.25) --
node[above, align=center] {%
$M_1:\; params\parallel ID_i\parallel N_{SGN_i}$\\
$epm\_pk_{SGN_i}\parallel\sigma^{(1)}_{SGN}$
}
(\xS,-1.25);

\draw[msgarrow] (\xS,-4.15) --
node[above, align=center] {%
$M_2:\; N_S\parallel epm\_pk_S\parallel ct_S$\\
$\parallel\sigma_S\parallel Finished_S$
}
(\xC,-4.15);

\draw[msgarrow] (\xC,-7.5) --
node[above, align=center] {%
$M_3:\; ct_{SGN_i}\parallel ECA$
}
(\xS,-7.5);

\end{tikzpicture}
\vspace{-3mm}
\caption{APQKE (Protocol~1): Ephemeral ML-KEM with long-term ML-DSA authentication, transcript binding ($TH_0,TH_1$), and Finished MACs, generating shared root key $RK_0$ for Algorithm~1.}
\label{fig:apqke-onecol}
\vspace{-5mm}
\end{figure*}

Fig.~\ref{fig:apqke-onecol} illustrates the APQKE handshake, which establishes an initial 256-bit $RK_0$ key for the DR channel. Unlike certificate-based designs, only long-term ML-DSA verification keys are provisioned out of band, as described in Section~\ref{sec:pki}; all ML-KEM public keys are ephemeral and exchanged during the handshake.

The $SGN_i$ initiates the handshake by executing steps (a)-(c). In step (a) it selects parameters $params$, identity $ID_i$, and a fresh nonce $N_{SGN_i}$, which is included in the transcript to provide freshness and replay resistance. In step (b), $SGN_i$ generates a one-time ML-KEM key pair $(epm\_sk_{SGN_i},epm\_pk_{SGN_i})$ that is erased after completion. In step (c), the ephemeral ML-KEM public key is authenticated using its long-term ML-DSA signing key. This early signature prevents an active adversary from substituting $epm\_pk_{SGN_i}$ during transmission because the server checks this signature before wrapping up the message. After signing, $SGN_i$ sends the message $M_1=(params,ID_i,N_{SGN_i},epm\_pk_{SGN_i}, \sigma^{(1)}_{SGN})$ to the server.

After receiving $M_1$, the server executes steps (a)--(i). In step (a), it uses $ID_i$ to fetch $pk^{sig}_{SGN_i}$ from its provisioned registry and verifies $\sigma^{(1)}_{SGN}$; if it fails, the handshake terminates before establishing any cryptographic state. Steps (b)--(c) contribute to server freshness by sampling $N_S$ and generating $(epm\_sk_S,epm\_pk_S)$. The server seeds the transcript in step (d) as $TH_0=H(params\parallel ID_i\parallel N_{SGN_i}\parallel N_S \parallel epm\_pk_{SGN_i}\parallel\sigma^{(1)}_{SGN})$, encapsulates it to the $SGN_i$ ephemeral key in step (e), obtaining $(ct_S,ss_S)$, and completes transcript binding in step (f) as $TH_1=H(TH_0\parallel epm\_pk_S\parallel ct_S)$.

The server authenticates the bound transcript by signing $TH_1$ (step (g)) and derives handshake traffic keys from $ss_S$ (step (h)): $hs=\mathrm{HKDF}(ss_S,\text{``hs''}\parallel TH_1)$ and $(hsK_{snd},hsK_{rcv})\leftarrow \mathrm{HKDF}(hs,\text{``hs-traffic''})$. This ordering is the server-side assignment. The $SGN_i$ derives the same two HKDF outputs but assigns them in reverse order as $(hsK_{rcv},hsK_{snd})\leftarrow \mathrm{HKDF}(hs,\text{``hs-traffic''})$. Hence, $hsK^{S}_{snd}=hsK^{SGN_i}_{rcv}$ and $hsK^{S}_{rcv}=hsK^{SGN_i}_{snd}$. Here, each endpoint considers $hsK_{snd}$ as the key for records it transmits and $hsK_{rcv}$ for records it receives. The server then computes a Finished MAC (step (i)) and sends $M_2=(N_S,epm\_pk_S,ct_S,\sigma_S,Finished_S)$ to the $SGN_i$.

On receipt of $M_2$, $SGN_i$ reconstructs $TH_0,TH_1$, decapsulates $ct_S$ to recover $ss_S$, derives $hs$, applies the SGN-side role assignment above, and verifies both the server signature $\sigma_S$ and $Finished_S\stackrel{?}{=} \mathrm{MAC}_{hsK_{rcv}} (\text{``server finished''}\parallel TH_1)$. At this point, $SGN_i$ accepts the server and confirms agreement on the transcript and traffic keys.

After authenticating the server, $SGN_i$ computes the second KEM secret by encapsulating $epm\_pk_S$ and generates $(ct_{SGN_i},ss_{SGN_i})$ (step (h)), and binds this ciphertext using $h_{SGN_i}=H(ct_{SGN_i})$ (step (i)). It then derives the initial Double-Ratchet root key (step (j)) $RK_0=\mathrm{HKDF} (ss_{SGN_i}\parallel ss_S,\text{``dr-root''}\parallel TH_1)$. Finally, it computes key confirmation $Finished_{SGN_i} =\mathrm{MAC}_{hsK_{snd}} (\text{``SGN finished''}\parallel TH_1\parallel h_{SGN_i})$ in step (k), encrypts it in step (l) as $ECA=\mathrm{AEAD.Enc}_{hsK_{snd}} (\text{AAD}=TH_1\parallel h_{SGN_i},Finished_{SGN_i})$, and sends $M_3=(ct_{SGN_i},ECA)$ to $S$.

When $S$ receives $M_3$, it computes $h_{SGN_i}=H(ct_{SGN_i})$, decrypts $ECA$ under $hsK_{rcv}$,
and verifies $Finished_{SGN_i}$. It then decapsulates $ct_{SGN_i}$ in constant time to obtain $ss_{SGN_i}$, derives the same $RK_0$, and erases ephemeral secrets. Both endpoints now hold the same $RK_0$, while the authenticated transcript hash $TH_1$ is available to bind the traffic of the DR protocol to this handshake.

%--------------------------------------------------------------------
\vspace{-2mm}
\subsection{Double-Ratcheted Data Channel}
\label{sec:proto2}
%\vspace{-2mm}

\begin{algorithm*}[!t]
\footnotesize
\centering
\caption{Session-aware Double Ratchet Channel (DH = X25519, AEAD = ChaCha20--Poly1305)}
\begin{tabular}{@{}p{0.98\linewidth}@{}}
\hline
\textbf{Shared constants:} $\kappa=256$; skip window $W$;
associated data $AD:=TH_1$. \\
\textbf{State (each endpoint):}
$RK,CK^{\mathsf{s}},CK^{\mathsf{r}}\in\{0,1\}^{\kappa}$;
counters $n_{\mathsf{s}},n_{\mathsf{r}},PN\in\mathbb{N}$; \\
\phantom{\textbf{State (each endpoint):}}
DH keypair $(sk,pk)$ and peer key $\widehat{pk}$;
skip-cache $\mathcal{K}_{\mathrm{skip}}$;
flag $\mathsf{needDH}\in\{0,1\}$.
The complete mutable state is denoted by $\mathcal{S}$. \\
\hline

\textbf{Bootstrap (run once after Protocol~1):}\\
1.\quad $RK\leftarrow RK_0$;
$n_{\mathsf{s}},n_{\mathsf{r}},PN\leftarrow 0$;
$\widehat{pk}\leftarrow \bot$;
$\mathcal{K}_{\mathrm{skip}}\leftarrow \emptyset$;
$\mathsf{needDH}\leftarrow 0$. \\

2.\quad $K_{\mathrm{boot}}
\leftarrow \mathrm{HKDF}
(RK_0,\texttt{"dr-boot"}\parallel AD,\kappa)$. \\

3.\quad \emph{SGN $\rightarrow$ S:}
$(sk,pk)\leftarrow \mathsf{DH.Gen}()$;
$\tau_C\leftarrow
\mathrm{MAC}_{K_{\mathrm{boot}}}
(\texttt{"SGN bootstrap"}\parallel AD\parallel pk)$;
send $B_0=(pk,\tau_C)$. \\

4.\quad \emph{S:}
verify
$\tau_C\stackrel{?}{=}
\mathrm{MAC}_{K_{\mathrm{boot}}}
(\texttt{"SGN bootstrap"}\parallel AD\parallel pk)$;
set $\widehat{pk}\leftarrow pk$;
$(sk,pk)\leftarrow \mathsf{DH.Gen}()$;
$\tau_S\leftarrow
\mathrm{MAC}_{K_{\mathrm{boot}}}
(\texttt{"server bootstrap"}\parallel AD\parallel pk)$;
send $B_1=(pk,\tau_S)$. \\

5.\quad \emph{SGN:}
verify
$\tau_S\stackrel{?}{=}
\mathrm{MAC}_{K_{\mathrm{boot}}}
(\texttt{"server bootstrap"}\parallel AD\parallel pk)$;
set $\widehat{pk}\leftarrow pk$. \\

6.\quad Both compute
$Z_0\leftarrow\mathsf{DH}(sk,\widehat{pk})$
and derive
$(RK,CK_0,CK_1)\leftarrow
\mathrm{HKDF}
(RK_0,\texttt{"dr-init"}\parallel Z_0\parallel AD,3\kappa)$. \\

7.\quad Role assignment:
SGN sets
$(CK^{\mathsf{s}},CK^{\mathsf{r}})
\leftarrow(CK_0,CK_1)$;
server sets
$(CK^{\mathsf{s}},CK^{\mathsf{r}})
\leftarrow(CK_1,CK_0)$.
Erase $K_{\mathrm{boot}}$. \\
\hline

\textbf{Header formats (bandwidth optimized):}\\
8.\quad Rekey header:
$\mathsf{HDR}_1=(1,pk,PN,n_{\mathsf{s}})$;
\quad Data header:
$\mathsf{HDR}_0=(0,n_{\mathsf{s}})$. \\
\hline

\textbf{StartSession():}
(called by the entity that sends first in a new session)\\
9.\quad $PN\leftarrow n_{\mathsf{s}}$;
$n_{\mathsf{s}}\leftarrow 0$;
$(sk,pk)\leftarrow\mathsf{DH.Gen}()$;
$Z\leftarrow\mathsf{DH}(sk,\widehat{pk})$;
$(RK,CK^{\mathsf{s}})
\leftarrow\mathrm{KDF}_{RK}(RK,Z)$;
$\mathsf{needDH}\leftarrow 1$. \\
\hline

\textbf{Send}$(m)$:\\
10.\quad If $\mathsf{needDH}=1$ then set
$hdr\leftarrow\mathsf{HDR}_1$ and
$\mathsf{needDH}\leftarrow 0$;
else set $hdr\leftarrow\mathsf{HDR}_0$. \\

11.\quad
$(CK^{\mathsf{s}},MK)
\leftarrow\mathrm{KDF}_{CK}(CK^{\mathsf{s}})$;
$nonce\leftarrow\mathrm{Encode96}(n_{\mathsf{s}})$. \\

12.\quad
$C\leftarrow
\mathsf{AEAD.Enc}
(MK,nonce,\mathrm{AAD}=AD\parallel hdr,m)$;
erase $MK$;
$n_{\mathsf{s}}\leftarrow n_{\mathsf{s}}+1$. \\

13.\quad Transmit $\mathsf{MSG}=(hdr,C)$. \\
\hline

\textbf{Receive}$(hdr,C)$:\\
14.\quad Parse $hdr$.
If $hdr=\mathsf{HDR}_1$, extract
$(pk^{peer},PN^{peer},n)$;
else extract $(n)$ and set
$pk^{peer}\leftarrow\widehat{pk}$. \\

15.\quad Set
$\mathcal{S}'\leftarrow\mathrm{Copy}(\mathcal{S})$;
operations in lines~16--19 are performed on
$\mathcal{S}'$.
If $hdr=\mathsf{HDR}_1$ and
$pk^{peer}\neq\widehat{pk}$, run
\textbf{DHRatchet}$(pk^{peer},PN^{peer})$
on $\mathcal{S}'$. \\

16.\quad If
$(\widehat{pk},n)\in\mathcal{K}_{\mathrm{skip}}$
then set
$MK\leftarrow
\mathcal{K}_{\mathrm{skip}}[\widehat{pk},n]$
and delete it from the temporary state. \\

17.\quad Else if
$n>n_{\mathsf{r}}+W$ then discard $\mathcal{S}'$,
drop and rate-limit the record, signal desynchronization, and resynchronize via an authenticated control path. \\

18.\quad Else if $n<n_{\mathsf{r}}$ then discard
$\mathcal{S}'$ and reject the record as a replay;
otherwise, while $n_{\mathsf{r}}\le n$ do
$(CK^{\mathsf{r}},MK)
\leftarrow\mathrm{KDF}_{CK}(CK^{\mathsf{r}})$;\\
\phantom{18.\quad Else if $n<n_{\mathsf{r}}$ then }
if $n_{\mathsf{r}}<n$ then store $MK$ as
$\mathcal{K}_{\mathrm{skip}}
[\widehat{pk},n_{\mathsf{r}}]$; \quad
\phantom{18.\quad Else if $n<n_{\mathsf{r}}$ then }
$n_{\mathsf{r}}\leftarrow n_{\mathsf{r}}+1$;
end while. \\

19.\quad
$nonce\leftarrow\mathrm{Encode96}(n)$;
attempt
$m\leftarrow
\mathsf{AEAD.Dec}
(MK,nonce,\mathrm{AAD}=AD\parallel hdr,C)$.
On failure, discard $\mathcal{S}'$, erase $MK$, and abort;
on success, commit
$\mathcal{S}\leftarrow\mathcal{S}'$,
erase $MK$, and output $m$. \\
\hline

\textbf{DHRatchet}$(pk^{peer},PN^{peer})$:\\
20.\quad Derive and cache at most $W$ skipped keys
from the current receive chain up to $PN^{peer}$;
abort the candidate ratchet if this bound is exceeded. \\

21.\quad $\widehat{pk}\leftarrow pk^{peer}$;
$n_{\mathsf{r}}\leftarrow 0$. \\

22.\quad $Z\leftarrow\mathsf{DH}(sk,\widehat{pk})$;
$(RK,CK^{\mathsf{r}})
\leftarrow\mathrm{KDF}_{RK}(RK,Z)$. \\

23.\quad $PN\leftarrow n_{\mathsf{s}}$;
$n_{\mathsf{s}}\leftarrow 0$;
$(sk,pk)\leftarrow\mathsf{DH.Gen}()$;
$Z'\leftarrow\mathsf{DH}(sk,\widehat{pk})$;
$(RK,CK^{\mathsf{s}})
\leftarrow\mathrm{KDF}_{RK}(RK,Z')$;
$\mathsf{needDH}\leftarrow 1$. \\
\hline
\end{tabular}
\label{alg:proto2}
\end{algorithm*}
%\vspace{-2mm}

The DR channel, outlined in Algorithm~\ref{alg:proto2} secures smart grid traffic after the APQKE handshake by evolving keys across messages and sessions with minimal steady-state overhead. It uses X25519 for both bootstrap and DH-ratchet updates, so initialization is post-quantum; however, recovery after ratchet-state disclosure relies on \emph{classical security assumptions} in order to minimize the overhead and reduce computational and bandwidth costs. The channel starts with the shared $RK_0$ and is linked to the authenticated handshake by setting $AD:=TH_1$ and using $AD$ as associated data in every authenticated encryption with associated data (AEAD) call. The ratcheted channel is initialized with two bootstrap messages, denoted \(B_0\) and \(B_1\).

Referring to Algorithm~\ref{alg:proto2}, \emph{line~1} initializes the local DR state from $RK_0$ (root/chain keys, counters, an unset peer DH key $\widehat{pk}$, an empty skip-cache $\mathcal{K}_{\mathrm{skip}}$, and $\mathsf{needDH}\leftarrow 0$). \emph{Line~2} derives a short-lived bootstrap authenticator $K_{\mathrm{boot}}$ from $(RK_0,AD)$. In \emph{line~3}, the SGN generates an initial DH keypair and sends $B_0=(pk,\tau_C)$, where $\tau_C$ binds the role label \texttt{"SGN bootstrap"}, $AD$, and the SGN public key under $K_{\mathrm{boot}}$. In \emph{line~4}, the server verifies this role-labeled MAC, stores the peer DH public key, generates its own DH keypair, and replies with $B_1=(pk,\tau_S)$, where $\tau_S$ uses the distinct label \texttt{"server bootstrap"}. After the SGN verifies $\tau_S$ (\emph{line~5}), both sides compute the initial DH shared secret $Z_0$ and derive the initial ratchet state $(RK,CK_0,CK_1)$ (\emph{line~6}). Because $K_{\mathrm{boot}}$ is derived from $RK_0$, verification of $B_0$ confirms to the server that the SGN derived $RK_0$, while verification of $B_1$ gives the SGN explicit confirmation that the server derived the same $RK_0$. Line~7 assigns the chain keys directionally (SGN uses $(CK^{\mathsf{s}},CK^{\mathsf{r}}) \leftarrow(CK_0,CK_1)$; the server swaps them) and erases $K_{\mathrm{boot}}$.

\emph{Line~8} defines two header formats. First is a compact data header $\mathsf{HDR}_0=(0,n_{\mathsf{s}})$ for most messages, and a rekey header $\mathsf{HDR}_1=(1,pk,PN,n_{\mathsf{s}})$ that includes the sender's DH public key and previous sending-chain length $PN$, which is used only for DH updates.

\texttt{StartSession()} is called by the entity that sends first in a new session. In \emph{line~9}, the sender snapshots the length of its previous sending chain ($PN\leftarrow n_{\mathsf{s}}$), resets $n_{\mathsf{s}}$, generates a fresh DH keypair, mixes a DH shared secret into the root state (updating $(RK,CK^{\mathsf{s}})$), and sets $\mathsf{needDH}\leftarrow 1$ so the next outbound message carries $\mathsf{HDR}_1$.

When sending an application message, \emph{line~10} selects the header: if $\mathsf{needDH}=1$, the sender transmits $\mathsf{HDR}_1$ once and clears $\mathsf{needDH}$; otherwise, it transmits $\mathsf{HDR}_0$. \emph{Line~11} advances the sending chain to obtain a fresh one-time message key $MK$ and computes the AEAD nonce from the current counter $n_{\mathsf{s}}$. \emph{Line~12} encrypts under AEAD with associated data $AD\parallel hdr$, then erases $MK$ and increments $n_{\mathsf{s}}$. \emph{Line~13} transmits $\mathsf{MSG}=(hdr,C)$.

Upon reception, \emph{line~14} parses the header to recover the message index $n$ and, in the case of $\mathsf{HDR}_1$, the peer DH public key and the peer's previous message number $PN^{peer}$; for $\mathsf{HDR}_0$, the receiver instead relies on its locally stored peer DH public key $\widehat{pk}$. \emph{Line~15} copies the complete receive state into a temporary candidate state $\mathcal{S}'$. If a rekey header contains a new peer DH public key, the \textbf{DHRatchet} procedure operates only on this candidate state. Therefore, an unauthenticated rekey header cannot immediately change the persistent root key, chain keys, counters, DH keys, or skip-cache.

\emph{Line~16} checks whether the message key is available in the candidate skip-cache, allowing out-of-order processing. If the message index exceeds the bounded skip window, \emph{line~17} discards the candidate state, drops and rate-limits the record, and permits resynchronization only through an authenticated control path; an unauthenticated header no longer triggers Protocol~1. If $n<n_{\mathsf{r}}$ and no corresponding skipped key exists, line~18 rejects the message as a replay. Otherwise, line~18 advances the candidate receiving chain to index $n$ and caches intermediate keys for skipped messages. Line~19 then performs authenticated decryption using associated data $AD\parallel hdr$. The candidate state is committed only after successful AEAD verification; on failure, it is discarded without changing the persistent ratchet state.

If \textbf{DHRatchet} is invoked, \emph{line~20} derives and caches at most $W$ remaining keys from the old receive chain up to $PN^{peer}$ and aborts the candidate ratchet if the bound is exceeded. \emph{Line~21} installs the new peer DH public key and resets the receive counter. \emph{Line~22} performs a DH computation with the current local DH secret key and the new peer public key to refresh $(RK,CK^{\mathsf{r}})$. \emph{Line~23} immediately prepares the next sending epoch by generating a fresh local DH keypair, refreshing $(RK,CK^{\mathsf{s}})$ again, and setting $\mathsf{needDH}\leftarrow 1$ so the next outbound message advertises the new DH public key using $\mathsf{HDR}_1$. When invoked during message reception, these updates remain temporary and are committed only after the associated ciphertext passes AEAD verification in \emph{line~19}.

%-------------------------------------------------------------------------------
\vspace{-2mm}
\section{Proposed Entropy Attack Suite (EAS)}
\label{sec:EAS}
\vspace{-1mm}
%-------------------------------------------------------------------------------
%We evaluated APQKE and Double Ratchet by simulating two entropy failures on a microcontroller: early-boot entropy holes and RNG path subversion, as shown in Figure \ref{fig:healthy-rng-critical}.

We evaluate the APQKE handshake under two SGN-side entropy-failure conditions and analyze their impact on the DR channel: an early-boot entropy hole and RNG-path subversion. Experiments ran on a Raspberry Pi Compute Module 5 as outlined in Section \ref{sec:sec-eval}. Fig.~\ref{fig:healthy-rng-critical} shows the injection points on the SGN side: $\mathcal{A}_{\mathrm{seed}}$ affects boot-time DRBG initialization, whereas $\mathcal{A}_{\mathrm{rng}}$ replaces bytes returned by the RNG API; server-side randomness remains healthy.

In our implementation, all SGN side randomness obtained by the APQKE handshake (SGN nonce, ML‑KEM ephemeral key generation, and encapsulation coins) is obtained using the SGN platform RNG interface shown in Fig.~\ref{fig:healthy-rng-critical} (operating system (OS) cryptographically secure pseudorandom number generator (CSPRNG)/ deterministic random bit generator (DRBG) seeded from a hardware entropy source). 

%The entropy generated by both an OS-only configuration and a TPM‑seeded configuration, in which the TPM is queried only for a small amount of seed material at session start.

% ---------------------- FIGURE 1 (vertical, half-column friendly) ----------------------
\tikzstyle{title} = [fill=white,font=\fontsize{8}{8}\selectfont\color{black!50}\ttfamily]
\tikzstyle{block} = [draw, fill=white, rectangle, align=center, minimum height=1em, minimum width=2em,
                     font=\fontsize{8}{7}\selectfont\color{black}\ttfamily]
\tikzstyle{smallblock} = [draw, fill=white, rectangle, align=center, minimum height=1em, minimum width=2em,
                          font=\fontsize{8}{7}\selectfont\color{black}\ttfamily]
\tikzstyle{specialblock} = [draw, fill=white, rectangle, align=center, minimum height=1em, minimum width=2em,
                            font=\fontsize{8}{7}\selectfont\color{black}\ttfamily]
\tikzstyle{logo} = [draw, fill=white, rectangle, minimum height=1em, minimum width=1em]
\definecolor{cBad}{RGB}{255,235,238}    % light red/pink
% NOTE: If you are not using the "stacked" style anywhere, remove it.
% If you keep it, you MUST load: \usetikzlibrary{shadows,arrows.meta}
% \tikzstyle{stacked} = [double copy shadow, shadow xshift=2pt, shadow yshift=-2pt,
%                        every edge/.append style = {draw, semithick, -Stealth}]

\begin{figure}[!htb]
\centering
\begin{tikzpicture}[auto, node distance=2.2cm, >=latex']

% -------------------- Top title --------------------
% \node[specialblock, fill=Grey, name=fg_title, yshift=-8pt, text width=6.6cm]
% {Entropy Seeding Point \\ {\fontsize{7}{8}\selectfont\textcolor{gray}{APQKE + Double Ratchet}}};
\node[coordinate, name=fg_title, yshift=-8pt] {};

% -------------------- Column headers (REMOVED blocks, keep anchors) --------------------
\node[coordinate, name=SGN,    below of=fg_title, yshift=40pt, xshift=-2.55cm] {};
\node[coordinate, name=server, below of=fg_title, yshift=40pt, xshift= 2.55cm] {};

% -------------------- SGN path --------------------
\node[block, name=c_trng, below of=SGN, yshift=28pt, text width=3.1cm]
{SGN: Hardware TRNG \\ {\fontsize{7}{8}\selectfont\textcolor{gray}{Physical noise source}}};

\node[block, name=c_seed, fill=cBad, below of=c_trng, yshift=20pt, text width=3.1cm]
{Entropy seeding interface \\ {\fontsize{7}{8}\selectfont\textcolor{gray}{Critical injection point}}};

\node[block, name=c_drbg, below of=c_seed, yshift=20pt, text width=3.1cm]
{CSPRNG / DRBG \\ {\fontsize{7}{8}\selectfont\textcolor{gray}{Seeded + reseeded}}};

\node[block, name=c_apqke, below of=c_drbg, yshift=16pt, text width=3.4cm]
{APQKE Handshake \\ {\fontsize{7}{8}\selectfont\textcolor{gray}{$N_{SGN_i}$, ephemerals, KEM coins}}};

\node[block, name=c_rk, below of=c_apqke, yshift=12pt, text width=3.1cm]
{Derived root key $RK_0$ \\ {\fontsize{7}{8}\selectfont\textcolor{gray}{Double Ratchet}}};

% -------------------- Server path --------------------
\node[block, name=s_trng, below of=server, yshift=28pt, text width=3.1cm]
{Server: Hardware TRNG \\ {\fontsize{7}{8}\selectfont\textcolor{gray}{Physical noise source}}};

\node[block, name=s_seed, below of=s_trng, fill=cBad, yshift=20pt, text width=3.1cm]
{Entropy seeding interface \\ {\fontsize{7}{8}\selectfont\textcolor{gray}{Healthy\\ RNG}}};

\node[block, name=s_drbg, below of=s_seed, yshift=20pt, text width=3.1cm]
{CSPRNG / DRBG \\ {\fontsize{7}{8}\selectfont\textcolor{gray}{Seeded + reseeded}}};

\node[block, name=s_apqke, below of=s_drbg, yshift=16pt, text width=3.1cm]
{APQKE Handshake \\ {\fontsize{7}{8}\selectfont\textcolor{gray}{$N_S$, ephemerals, KEM coins}}};

\node[block, name=s_rk, below of=s_apqke, yshift=12pt, text width=3.1cm]
{Derived root key $RK_0$ \\ {\fontsize{7}{8}\selectfont\textcolor{gray}{Double Ratchet}}};

% -------------------- Vertical arrows --------------------
% -------------------- Vertical arrows --------------------
% \draw[thick, ->] (SGN) -- (c_trng);      % removed (header removed)
\draw[thick, ->] (c_trng) -- (c_seed);
\draw[thick, ->] (c_seed) -- (c_drbg);
\draw[thick, ->] (c_drbg) -- (c_apqke);
\draw[thick, ->] (c_apqke) -- (c_rk);

% \draw[thick, ->] (server) -- (s_trng);  % removed (header removed)
\draw[thick, ->] (s_trng) -- (s_seed);
\draw[thick, ->] (s_seed) -- (s_drbg);
\draw[thick, ->] (s_drbg) -- (s_apqke);
\draw[thick, ->] (s_apqke) -- (s_rk);

% -------------------- Dashed highlight --------------------
\node[draw, dashed, thick, color=gray, inner xsep=4pt, inner ysep=5pt,
      fit=(c_seed)] (c_seed_hl) {};
\node[draw, dashed, thick, color=gray, inner xsep=4pt, inner ysep=5pt,
      fit=(s_seed)] (s_seed_hl) {};

\draw[draw=none] (c_seed_hl.east) --
  node[midway, title, yshift=5pt, align=center]
  {\textbf{\shortstack{Critical\\injection\\point}}}
  (s_seed_hl.west);

% -------------------- Handshake message flow --------------------
\draw[thick, <->]
(c_apqke.east) --
node[pos=0.5, above, fill=none, title, yshift=13.05pt, align=center]
{M1: params, $ID_i$, $N_{SGN_i}$,\\ $\mathsf{epm}_{SGN_i}$, $\sigma_{SGN}$}
node[pos=0.5, below, title, yshift=-13.05pt, align=center]
{M2: $N_S$, $\mathsf{epm}_S$, $ct_S$, $\sigma_S$, $Finished_S$\\
 M3: $ct_{SGN_i}$, $ECA$}
(s_apqke.west);

\end{tikzpicture}
\vspace{-6mm}
\caption{Entropy seeding points in APQKE and the Double Ratchet (message fields aligned with APQKE protocol).}
\label{fig:healthy-rng-critical}
\vspace{-6mm}
\end{figure}
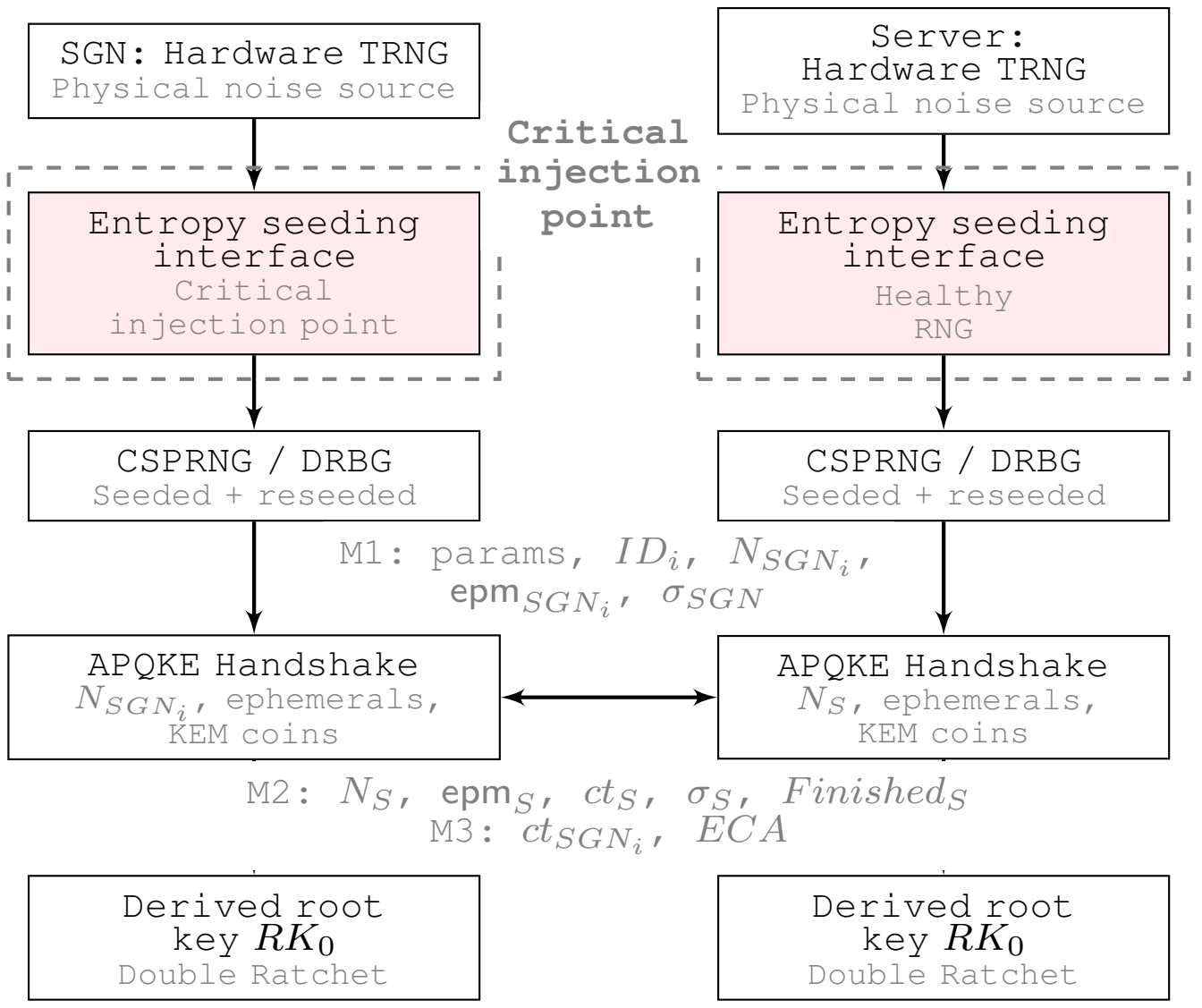

\vspace{-3mm}
\subsection{Early-boot Entropy Hole}
\label{sec:eas-boot}
\vspace{-2mm}

Smart-grid nodes may reset after outages, maintenance, or updates, creating an early-boot entropy risk if cryptographic operations begin before the endpoint DRBG/CSPRNG is securely initialized \cite{nistsp80090b}. Whether this risk arises depends on the application's randomness interface. On Linux, \texttt{getrandom()} with zero flags waits for RNG initialization, whereas direct reads from \texttt{/dev/urandom} may return output earlier \cite{muller2024entropy}. Seeding an application-level generator from a timestamp can also leave its output predictable if an attacker knows or controls the approximate reset time \cite{eastlake2005randomness}.

The APQKE handshake protocol requires device-side randomness for the SGN nonce, ephemeral ML-KEM key generation, and the encapsulation coins. Predictability in these inputs can expose the handshake root key $RK_0$. To evaluate this failure mode in a controlled way, we implemented an \emph{early-boot RNG mode} at the entropy boundary (i.e., we intercept the seeding interface in Fig.~\ref{fig:healthy-rng-critical}). The weak-RNG interval is injected into the experimental platform\footnote{Raspberry Pi Compute Module~5 serves as the experimental platform} and does not represent a measured failure of its normal Linux RNG.

The model is parameterized by (i) an injected weak-RNG interval $T_{\mathrm{inj}}$ and (ii) an effective early-boot seed space of size $2^b$ (where $b$ is the number of unknown early-boot seed bits). During $[0,T_{\mathrm{inj}})$, the RNG output stream is deterministic given a boot seed $s\in\{0,1\}^b$; after $T_{\mathrm{inj}}$, the RNG behaves as a healthy source. Formally, letting $\mathsf{RNG}(\cdot)$\footnote{We start APQKE after a configurable delay $d$ from boot and measure whether the attacker recovers $RK_0$. The notation $\mathsf{RNG}(\cdot)$ refers to the device RNG interface.} output bytes and $G$ denote the device DRBG/PRG, where $U(\{0,1\}^{8})$ denotes an idealized uniform random byte.

\begin{equation}\label{eq:1}
\mathrm{RNG}(t)=
\begin{cases}
G(s,t), & 0\leq t<T_{\mathrm{inj}},\\
U(\{0,1\}^{8}), & t\geq T_{\mathrm{inj}}.
\end{cases}
\end{equation}

\noindent \textbf{Boot-Window Seed Recovery Attack (BWSR):} Fig.~\ref{fig:entropy_boot_warmup} shows how an attacker exploits the injected weak-RNG interval. The attacker records an APQKE handshake, enumerates possible boot seeds, and compares the regenerated handshake values with the transcript to identify a seed that permits reconstruction of $RK_0$. The attacker does not break ML-KEM or ML-DSA; it exploits a small effective seed space during the early-boot window.

\definecolor{cSGN}{RGB}{255,244,230} % light orange
\definecolor{cServer}{RGB}{232,245,233} % light green
\definecolor{cProto}{RGB}{232,240,252}  % light blue
\definecolor{cKey}{RGB}{255,249,196}    % light yellow
\definecolor{cAttack}{RGB}{245,245,245} % light gray
\definecolor{cBad}{RGB}{255,235,238}    % light red/pink

\begin{figure*}[!htb]
\centering
%\small
\def\workflowWidth{5cm}
\def\stackWidth{4.05cm}
\begin{tikzpicture}[
  workflow/.style={block,text width=\workflowWidth},
  font=\small,
  >=Latex,
  node distance=2cm,
  box/.style={draw, rectangle, font=\fontsize{8}{8}\color{black}\ttfamily, align=center, inner sep=5pt,
              minimum height=1em, minimum width=1em, fill=white},
  boxS/.style={draw, rectangle, font=\fontsize{8}{8}\color{black}\ttfamily, align=center,
               minimum height=1em, minimum width=1em, fill=white},
  bad/.style={draw=red!65!black, fill=cBad, minimum height=4em, minimum width=1em, rectangle, font=\fontsize{8}{8}\color{black}\ttfamily, align=center, dashed},
  pill/.style={draw=red!65!black, fill=red!12, rounded corners=10pt, align=center, inner sep=3pt},
  note/.style={font=\scriptsize, align=left},
  arrow/.style={->, thick},
  darrow/.style={->, thick, dashed},
]

% -------------------- Top-left: attack description --------------------
\node[specialblock, name=bwsr, fill=cBad, text width=50mm, minimum height=10mm]
{Boot-Window Seed Recovery Attack (BWSR) \\
 {\fontsize{7}{8}\selectfont\textcolor{gray}{Injected weak-RNG model Effective seed space $2^b$ Deterministic for $t < T_{inj}$}}};

% -------------------- Top-right: early-boot protocol stacks --------------------
% (Placement preserves the original coordinate deltas using node distance + x/y shifts)
\node[box, fill=cSGN, name=sm, right of=bwsr, xshift=3.45cm, yshift=0.025cm, inner xsep=.3333em, text width=\stackWidth]
{SGN \\
 {\fontsize{7}{8}\selectfont\textcolor{gray}{Early-boot RNG mode}}};

\node[specialblock, fill=cServer, name=scada, right of=sm, xshift=4cm, text width=\stackWidth]
{Server \\
 {\fontsize{7}{8}\selectfont\textcolor{gray}{Healthy RNG}}};

\node[block, name=pcli, below of=sm, yshift=1.05cm, text width=\stackWidth]
{APQKE + Double Ratchet};

\node[block, name=psrv, below of=scada, yshift=1.05cm, text width=\stackWidth]
{APQKE + Double Ratchet};

\node[block, name=rkcliTop, below of=pcli, yshift=1.20cm, text width=\stackWidth]
{Derived root key $RK_0$};

\node[block, name=rksrvTop, below of=psrv, yshift=1.20cm, text width=\stackWidth]
{Derived root key $RK_0$};

% Vertical flows
\draw[thick, ->] (sm) -- (pcli);
\draw[thick, ->] (scada) -- (psrv);
\draw[thick, ->] (pcli) -- (rkcliTop);
\draw[thick, ->] (psrv) -- (rksrvTop);

% Dashed "protocol exchange" hint
\draw[dashed, thick, color=gray]
  (pcli.south) -- ++(0,-0.15) -- ($(psrv.south)+(0,-0.15)$) -- (psrv.south);

% Same root key
\draw[<->, thick] (rkcliTop.east) -- (rksrvTop.west);

% BWSR influence arrow into the SGN early-boot path
\coordinate (SGNMid) at ($(sm.south)!0.55!(pcli.north)$);
\draw[thick, ->, dashed, color=gray] (bwsr.east) -- (sm.west);

% Dashed frame around the early-boot mode region (replacement for static InnerSW/InnerNE rectangle)
\node[draw, dashed, thick, color=gray, inner xsep=5pt, inner ysep=5pt,
      fit=(sm)(scada)(pcli)(psrv)(rkcliTop)(rksrvTop)] (TopFrame) {};

% -------------------- Bottom: measurement / evaluation workflow --------------------
\node[workflow, name=online, below of=bwsr, yshift=-0.76cm]
{Online (passive) \\
 {\fontsize{7}{8}\selectfont\textcolor{gray}{Record APQKE transcript\\
 ($M_1$, $M_2$, $M_3$)}}};

\draw[thick, ->] (bwsr.south) -- (online.north);

\node[workflow, name=seedsearch, right of=online, xshift=3.6cm]
{Offline seed search \\
 {\fontsize{7}{8}\selectfont\textcolor{gray}{Brute force $2^{b}$ boot seeds simulate RNG stream match $N_{SGN_i}$ prefix}}};

\node[workflow, name=replay, right of=seedsearch, xshift=3.6cm]
{Offline replay \\
 {\fontsize{7}{8}\selectfont\textcolor{gray}{Replay APQKE using the recovered RNG stream and recorded transcript}}};

\draw[thick, ->] (online.east) -- (seedsearch.west);
\draw[thick, ->] (seedsearch.east) -- (replay.west);

\node[block, name=measure, below of=online, yshift=0.55cm, text width=4.0cm]
{Measure $p_{\mathrm{KR}}(d)$ \\
 {\fontsize{7}{8}\selectfont\textcolor{gray}{Repeat sessions with start delay $d$; count $RK_0$ recoveries}}};

\node[block, name=warmup, right of=measure, xshift=3.5cm, text width=5.0cm]
{Observed threshold \\
 {\fontsize{7}{8}\selectfont\textcolor{gray}{Observed threshold $T_{0.5}^{inj} = min\{d : pKR(d) < 0.5\}$}}};

% Online -> Measure, Measure -> Warmup
\draw[thick, ->] (online.south) -- (measure.north);
\draw[thick, ->] (measure.east) -- (warmup.west);

% Replay -> Warmup (down then left into Warmup's north side)
\draw[thick, ->]
  (replay.south)
  -- ++(0,-0.13) coordinate (replayTurn)
  -- (warmup.north |- replayTurn)
  -- (warmup.north);

% Bottom recovered keys
\node[block, name=rkcliBot, fill=cKey, below of=measure, yshift=0.6cm, text width=3.1cm]
{$RK_0$ \\
 {\fontsize{7}{8}\selectfont\textcolor{gray}{SGN}}};

\node[block, name=rksrvBot, fill=cKey, below of=warmup, yshift=0.6cm, text width=3.1cm]
{$RK_0$ \\
 {\fontsize{7}{8}\selectfont\textcolor{gray}{Server}}};

\draw[<->, thick] (rkcliBot.east) -- (rksrvBot.west);

% Dashed bracket from (Measure, Warmup) to the recovered keys
\coordinate (brL) at ($(measure.south)+(0,-0.25)$);
\coordinate (brR) at ($(warmup.south |- brL)$); % same y as brL

\draw[dashed, thick, color=gray] (measure.south) -- (brL);
\draw[dashed, thick, color=gray] (warmup.south) -- (brR);
\draw[dashed, thick, color=gray] (brL) -- (brR);
\draw[thick, ->] (brL) -- (rkcliBot.north);
\draw[thick, ->] (brR) -- (rksrvBot.north);

% -------------------- Right: PKR vs delay plot --------------------
\node[block, name=plot, draw=none, right of=warmup, xshift=4.2cm, yshift=-0.80cm,
      minimum width=6.85cm, minimum height=3.0cm] {};

\begin{scope}[shift={($(plot.south west)+(0.75,0.60)$)}]
  \draw[->, thick] (0,0) -- (0,1.9) node[above] {$p_{\mathrm{KR}}$};
  \draw[->, thick] (0,0) -- (5.1,0) node[right] {$d(\mathrm{ms})$};

  % tick at 1
  \node[left] at (0,1.45) {1};
  \draw[dashed, color=gray] (0,1.45) -- (0.45,1.45);

  % stylized step curve
  \draw[thick]
    (0.00,0.25) -- (0.00,1.45)
    -- (3.15,1.45)
    -- (3.15,0.55)
    -- (3.75,0.55)
    -- (3.75,0.25)
    -- (5.00,0.25);

  % Twarm marker
  \draw[dashed, color=gray] (3.75,0.25) -- (3.75,1.45);
  \draw[dashed, color=gray] (3.75,1.45) -- (-0.10,1.45);
  \node[right] at (3.75,1.05) {$T_{\mathrm{inj}}$};
\end{scope}

\end{tikzpicture}
\vspace{-8mm}
%\caption{Boot-Window Seed Recovery Attack.}
\caption{Boot-Window Seed Recovery Attack (BWSR): a passive attacker records an APQKE transcript, searches the $2^b$ early-boot seeds offline, and replays the handshake to recover $RK_0$.}
\label{fig:entropy_boot_warmup}
\vspace{-7mm}
\end{figure*}
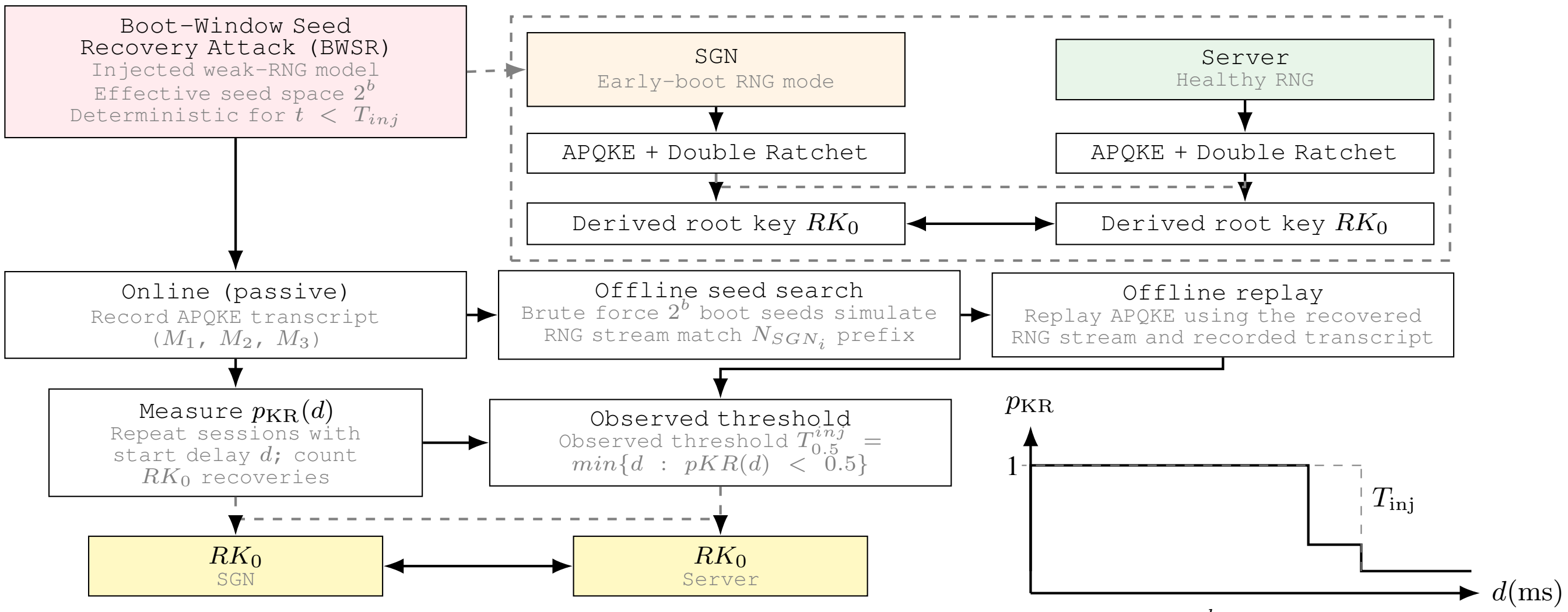

\emph{1) Online transcript recording:}
The attacker passively records an APQKE protocol run initiated at a delay $d$ after boot, observing $(M_1,M_2,M_3)$. In particular, it learns the public SGN nonce $N_{SGN_i}$ and the SGN's ephemeral KEM public key in $M_1$, and later observes $(epm\_pk_S,ct_S)$ in $M_2$ and the SGN's ciphertext $ct_{SGN_i}$ in $M_3$.

\emph{2) Offline boot-seed search:}
Because $(N_{SGN_i}, \mathit{epm\_pk}_{SGN_i})$ are derived from the early‐boot RNG stream, the adversary can mount an offline search over candidate boot seeds $s \in \{0,1\}^b$. For each candidate seed, the adversary simulates the early‐boot RNG output and regenerates the corresponding values $(N'_{SGN_i}, \mathit{epm\_pk}'_{SGN_i}, \mathit{epm\_sk}'_{SGN_i}) \gets \mathsf{SimRNG}(s)$, which are then validated against the values recorded during the APQKE handshake. A seed is accepted if the regenerated public values are consistent with the observed transcript; in particular, a prefix match on $N_{SGN_i}$ serves as an efficient filter to eliminate most incorrect candidates.

\emph{3) Offline APQKE protocol replay and $RK_0$ reconstruction:}
Using any seed consistent with the transcript, the attacker reproduces the remaining device-side RNG draws used by APQKE protocol (including the KEM key generation and encapsulation coins) and reconstructs the two KEM shared secrets:
(i) decapsulate the recorded $ct_S$ using $epm\_sk'_{SGN_i}$ to obtain $ss_S$, and
(ii) recompute the SGN encapsulation toward the recorded $epm\_pk_S$ to obtain $ss_{SGN_i}$ (and verify correctness by matching the resulting $ct_{SGN_i}$ to the recorded ciphertext).
Since $TH_1$ is a deterministic function of the public handshake values, the attacker reconstructs $TH_1$ and derives the same session root key \(RK_0 \;=\; \mathrm{HKDF}\!\left(ss_{SGN_i} \parallel ss_S,\ \texttt{"dr-root"} \parallel TH_1\right).\)

This measures attack success with $p_{\mathrm{KR}}(d)$, defined as the fraction of sessions started at delay $d$ after boot in which the attacker recovers $RK_0$. We define the characteristic time $T_{0.5}^{\mathrm{inj}} = \min\{d : p_{\mathrm{KR}}(d) < 0.5\}$, which is the smallest delay $d$ at which the attacker’s success probability in recovering $RK_0$ falls below $0.5$.

\vspace{-2mm}
\subsection{RNG Path Subversion}
\vspace{-1mm}

On the SGN, the handshake needs cryptographically secure random bytes to generate its nonce and ephemeral ML-KEM keys. These bytes come from the OS RNG API, ultimately seeded by a hardware entropy source. We refer to the chain from entropy source to OS generator to handshake code as the \emph{RNG path}. The \emph{early-boot entropy hole} is a transient fault at the start of this path: shortly after reset, the generator may be seeded with low entropy, making its output feasibly enumerable. \emph{RNG-path subversion} is a persistent fault near the end of the path: compromised firmware or software intercepts or replaces RNG API outputs with attacker-predictable bytes, regardless of the true entropy source. RNG subversion is a well-established threat model, previously analyzed in the context of TLS implementations \cite{checkoway2014practical} and noted in the Double Ratchet specification \cite{perrin2016double}. Below describes how an adversary who controls the RNG API output recovers the root key $RK_0$ from a recorded handshake, and what this recovery implies for the confidentiality of later Double Ratchet message keys. Section~\ref{subsec:EAS_results} reports the experimental results, including the mitigation methods.

%, and whether TPM-based seeding or measured-boot gating can mitigate the attack.

In the reported experiment, the constrained attacker $\mathcal{A}_{\mathrm{rng}}$ controls only the returned random bytes. The upstream entropy sources, server RNG, provisioned keys, and other process memory remain uncompromised. The controlled APQKE retrieves are the SGN nonce $N_{SGN_i}$, ephemeral ML-KEM key-generation randomness, and encapsulation coins.

We instantiate the malicious provider as a diagnostic 16-bit linear congruential generator (LCG). This deliberately weak generator provides a reproducible failure condition for the experiments. Starting from $\mathsf{st}_0$, it updates and outputs
\begin{align}
  \mathsf{st}_{j+1} &= (25173\mathsf{st}_j+13849)\bmod 2^{16}, \label{eq:lcg-state}\\
  r_j &= \mathsf{st}_{j+1}\bmod 2^8. \label{eq:lcg-lowbyte}
\end{align}
The RNG API returns the byte stream $(r_0,r_1,\ldots)$ in place of healthy randomness. Because $25173\bmod 2^8=85$ and $13849\bmod 2^8=25$, the observable bytes satisfy $r_{j+1}=(85r_j+25)\bmod 2^8.$

%Because $25$ is odd and $85\equiv1\pmod 4$, this low-byte recurrence has full period $2^8$. Consequently, the $2^8$ possible initial low-byte values define $2^8$ observable phases of the same period-$256$ sequence. Although the full LCG maintains $2^{16}$ possible internal states, each observable phase corresponds to an equivalence class of $2^8$ states that share the same initial low byte and produce the same protocol-visible sequence. The attack, therefore, identifies an observable stream phase rather than a unique 16-bit internal state.

%The attacker supplies the generator in Eqs.~\eqref{eq:lcg-state}--\eqref{eq:lcg-output}, so it knows the constants, and it knows the order in which the handshake requests random bytes. With both known, the attack needs one recorded handshake and no further access to the device. It has three steps (Fig.~\ref{fig:rng-subversion}): record the handshake, identify which of the $2^8$ observable streams the SGN used, and replay the handshake offline to recompute $RK_0$.

Because $25$ is odd and $85\equiv1\pmod 4$, the low-byte recurrence has full period $2^8$, so its $2^8$ possible initial values define distinct phases of the same sequence. The full LCG has $2^{16}$ internal states, but the $2^8$ states sharing a given initial low byte produce the same protocol-visible sequence. The attack therefore identifies an observable phase rather than a unique 16-bit state. The attacker supplies the generator in Eqs.~\eqref{eq:lcg-state}--\eqref{eq:lcg-output} and knows both its constants and the order in which the handshake requests random bytes. With this knowledge, the attack proceeds in three steps (Fig.~\ref{fig:rng_path_subversion_one}): record the handshake, identify the phase used by the SGN, and replay the handshake offline to recompute $RK_0$, without further device access.

\emph{1) Record the transcript:}
The attacker records $(M_1,M_2,M_3)$ and extracts $N_{SGN_i}$, the ephemeral ML-KEM public values, $ct_S$, and $ct_{SGN_i}$.

\emph{2) Stream recovery:}
The attacker compares the first byte of $N_{SGN_i}$ with the first nonce byte of each of the $2^8$ observable streams. Since Eq.~\eqref{eq:lcg-lowbyte} visits every byte value once per period, exactly one stream matches. The attacker confirms the match by regenerating the SGN's ephemeral ML-KEM key pair and comparing its public key with the one in $M_1$.

\emph{3) Offline replay:}
The attacker decapsulates $ct_S$ using the regenerated secret key to obtain $ss_S$. It then encapsulates to $epm\_pk_S$ with coins drawn from the same stream to obtain $ss_{SGN_i}$, and checks that the resulting ciphertext equals $ct_{SGN_i}$. It computes $TH_1$ from $M_1$ and $M_2$ and derives $RK_0 = \operatorname{HKDF}\left(ss_{SGN_i}\parallel ss_S,\,\texttt{"dr-root"}\parallel TH_1\right)$. Both shared secrets are recoverable from SGN-side randomness and the recorded transcript, so healthy server randomness does not protect $RK_0$. 

Recovery of $RK_0$ permits forgery of the $K_{\mathrm{boot}}$ authenticators and active interference with DR bootstrap. It does not, by itself, reveal chain keys to a passive classical attacker from an already completed honest bootstrap because initialization also mixes the fresh X25519 secret $Z_0$, which remains unknown to that attacker. If the same RNG subversion also controls X25519 bootstrap or ratchet key generation, the corresponding DH secrets become predictable, and classical post-compromise recovery remains unavailable, while the attacker can predict secrets introduced by subsequent ratchet updates.

The experiments evaluate mitigation of the RNG-path subversion attack and distinguish between insufficient seed entropy and substitution of RNG outputs after seeding. While TPM-based seeding injects entropy into the RNG, it cannot prevent software from replacing its outputs. In our evaluation, the enforced PCR policy prevented channel establishment when using the assessed modified firmware. Analogous evaluations can be conducted for additional protocols by adapting the attack implementation to each protocol’s sequence of RNG invocations, handshake message flow, and key-derivation steps.

% Optional: a simple, print-friendly palette (define once)
\definecolor{cSGN}{RGB}{255,244,230} % light orange
\definecolor{cServer}{RGB}{232,245,233} % light green
\definecolor{cProto}{RGB}{232,240,252}  % light blue
\definecolor{cKey}{RGB}{255,249,196}    % light yellow
\definecolor{cAttack}{RGB}{245,245,245} % light gray
\definecolor{cBad}{RGB}{255,235,238}    % light red/pink

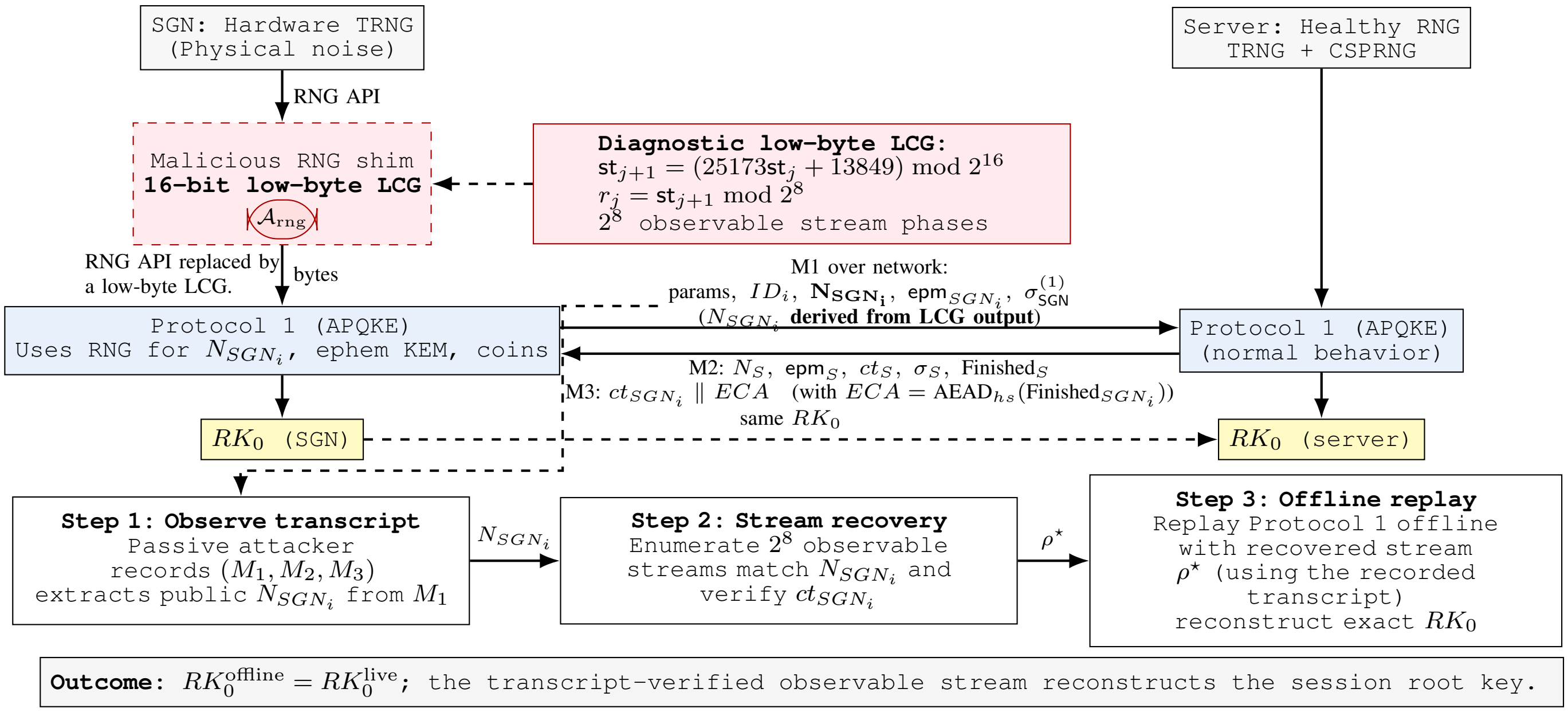
\begin{figure*}[!t]
\centering
\begin{tikzpicture}[
  font=\small,
  >=Latex,
  node distance=6mm,
  box/.style={draw, rectangle, font=\fontsize{8}{8}\color{black}\ttfamily, align=center, inner sep=5pt,
              minimum height=1em, minimum width=1em, fill=white},
  boxS/.style={draw, rectangle, font=\fontsize{8}{8}\color{black}\ttfamily, align=center,
               minimum height=1em, minimum width=1em, fill=white},
  bad/.style={draw=red!65!black, fill=cBad, minimum height=4em, minimum width=1em, rectangle, font=\fontsize{8}{8}\color{black}\ttfamily, align=center, dashed},
  pill/.style={draw=red!65!black, fill=red!12, rounded corners=10pt, align=center, inner sep=3pt},
  note/.style={font=\scriptsize, align=left},
  arrow/.style={->, thick},
  darrow/.style={->, thick, dashed},
]

% ---------------------------------------------------------
% Layout constants (easy tuning knobs)
% ---------------------------------------------------------
\def\Xsep{12.0}     % horizontal separation between SGN/server columns (cm)
\def\GapShimP{12mm} % gap between shim and Protocol 1 box

% =========================================================
% TOP: SGN stack (left)
% =========================================================
%\node[box, fill=cSGN] (Cdev) at (0.0,4.5) {SGN APQKE};

\node[boxS, fill=gray!6] (Ctrng) at (0.0,4.5)% [below=3mm of Cdev]
  {SGN: Hardware TRNG\\(Physical noise)};

\node[bad] (Cshim) [below=6mm of Ctrng]
  {Malicious RNG shim\\ \textbf{16-bit low-byte LCG} \\};

\node[boxS, fill=cProto] (Capqke) [below=7mm of Cshim]
  {Protocol 1 (APQKE)\\Uses RNG for $N_{SGN_i}$, ephem KEM, coins};

\node[boxS, fill=cKey] (RKc) [below=5mm of Capqke]
  {$RK_0$ (SGN)};

% SGN internal arrows
\draw[arrow] (Ctrng) -- node[right, font=\scriptsize] {RNG API} (Cshim);
\draw[arrow] (Cshim) -- node[right, font=\scriptsize] {bytes} (Capqke);
\draw[arrow] (Capqke) -- (RKc);

% Firmware attacker pill INSIDE the dashed shim box
\node[pill, font=\scriptsize] (fwatt) at ($(Cshim.south)+(0,3.0mm)$) {$\mathcal{A}_{\mathrm{rng}}$};
\node[note] (fwnote) at ($(Cshim.south)+(-1.15,-3.5mm)$)
{RNG API replaced by\\ a low-byte LCG.};

% LCG equation callout
\node[
  boxS,
  fill=cBad,
  draw=red!65!black,
  align=left,
  minimum width=62mm
] (LCGeq) at ($(Cshim)+(6.0,0)$) {
\textbf{Diagnostic low-byte LCG:}\\
$\mathsf{st}_{j+1}
=(25173\mathsf{st}_{j}+13849)\bmod 2^{16}$\\
$r_j=\mathsf{st}_{j+1}\bmod 2^8$\\
$2^8$ observable stream phases
};
\draw[darrow] (LCGeq.west) -- (Cshim.east);

% =========================================================
% TOP: Server stack (right)
% =========================================================
%\node[box, fill=cServer]  (Sdev)   at ($(Cdev)+(\Xsep,0)$)  {Server APQKE};
\node[boxS, fill=gray!6]  (Strng)  at ($(Ctrng)+(\Xsep,0)$) {Server: Healthy RNG\\TRNG + CSPRNG};
\node[boxS, fill=cProto]  (Sapqke) at ($(Capqke)+(\Xsep,0)$){Protocol 1 (APQKE)\\(normal behavior)};
\node[boxS, fill=cKey]    (RKs)    at ($(RKc)+(\Xsep,0)$)   {$RK_0$ (server)};

\draw[arrow] (Strng)  -- (Sapqke);
\draw[arrow] (Sapqke) -- (RKs);

% Same RK0 (dashed)
\draw[darrow] (RKc.east) -- node[above, font=\scriptsize] {same $RK_0$} (RKs.west);

% =========================================================
% Network messages (two parallel lines)
% =========================================================
\coordinate (m1L) at ($(Capqke.east)+(0.0, 0.15)$);
\coordinate (m1R) at ($(Sapqke.west)+(-0.0,0.15)$);
\coordinate (m2L) at ($(Sapqke.west)+(-0.0,-0.15)$);
\coordinate (m2R) at ($(Capqke.east)+(0.0,-0.15)$);

\draw[arrow] (m1L) -- (m1R);
\node[font=\scriptsize, align=center] (m1Lab) at ($(m1L)!0.5!(m1R)$)
{M1 over network:\\
$\text{params},\ ID_i,\ \mathbf{N_{SGN_i}},\ \mathsf{epm}_{SGN_i},\ \sigma^{(1)}_{\mathsf{SGN}}$\\
(\textbf{$N_{SGN_i}$ derived from LCG output})\\ \\ \\};

\draw[arrow] (m2L) -- (m2R);
\node[font=\scriptsize, align=center] (m2Lab) at ($(m2L)!0.5!(m2R)$)
{\\ \\ \\
M2: $N_S,\ \mathsf{epm}_S,\ ct_S,\ \sigma_S,\ \text{Finished}_S$\\
M3: $ct_{SGN_i}\ \|\ ECA$ \ \ (with $ECA=\text{AEAD}_{hs}(\text{Finished}_{SGN_i})$)};

% =========================================================
% Bottom attacker workflow
% =========================================================
\node (mid) at ($(RKc)!0.5!(RKs)$) {}; % anchor

\node[box, text width = 14em] (A1) at ($(mid)+(-6.475,-1.4)$) {
\textbf{Step 1: Observe transcript}\\
Passive attacker records $(M_1,M_2,M_3)$\\
extracts public $N_{SGN_i}$ from $M_1$
};

\node[box, text width = 14em] (A2) at ($(mid)+(-0.15,-1.4)$) {
\textbf{Step 2: Stream recovery}\\
Enumerate $2^8$ observable streams
match $N_{SGN_i}$ and\\ verify $ct_{SGN_i}$
};

\node[box, text width = 14.5em] (A3) at ($(mid)+(6.05,-1.4)$) {
\textbf{Step 3: Offline replay}\\
Replay Protocol~1 offline with 
recovered stream $\rho^\star$ (using the recorded\\ 
transcript) reconstruct exact $RK_0$
};

\draw[arrow] (A1) -- node[above, font=\scriptsize] {$N_{SGN_i}$} (A2);
\draw[arrow] (A2) -- node[above, font=\scriptsize] {$\rho^\star$} (A3);

\node[boxS, fill=cAttack, align=left, minimum width=125mm] (A4) at ($(mid)+(0.0,-2.8)$) {
\textbf{Outcome:} $RK_0^{\mathrm{offline}}=RK_0^{\mathrm{live}}$; the transcript-verified observable stream reconstructs the session root key.
};

% Connect Step 1 to the M1 label
\draw[darrow]
    ($(m1Lab.west)+(0,0.25)$)
  -- ++(-1.1,0.0)
  -- ++(0.0,-1.90)
  -- ++(-3.72,0)
  -- (A1.north);

% Outer frame
% \node[draw=gray!70, rounded corners=3pt, fit=(Cdev)(Sdev)(RKc)(RKs)(A1)(A3)(A4)(fwnote),
%       inner sep=7pt] {};

\end{tikzpicture}
\vspace{-3mm}
\caption{RNG-path subversion and offline root-key recovery under the diagnostic 16-bit low-byte LCG}
\label{fig:rng_path_subversion_one}
\vspace{-7mm}
\end{figure*}

%===========================================
% Results and Analysis
%===========================================
\vspace{-3mm}
\section{Results and Analysis}
\label{sec:sec-eval}
\vspace{-2mm}
This section evaluates APQKE-DR under the threat model of Section~\ref{sec:threat}. We first present the security analysis results in Section \ref{subsec:secuirty}, then separately report performance measurements for APQKE and DR, and finally discuss the EAS results and their deployment implications. For this purpose, all experiments\footnote{The server ran Ubuntu 22.04.5 on an Intel Core i9-14900 (32 cores) with 32 GB RAM. The SGN endpoint used a Raspberry Pi Compute Module 5 (Broadcom BCM2712 quad-core 64-bit Arm Cortex-A76 at 2.4 GHz) with Gigabit Ethernet. We connected a TPM 9670 via SPI as a hardware entropy source for entropy-failure experiments.} were conducted on a two-node client/server testbed that simulates a utility head-end communicating with a field SGN.

\vspace{-3mm}
\subsection{ProVerif: Security Analysis}\label{subsec:secuirty}
\vspace{-2mm}
We evaluated APQKE protocol in ProVerif~\cite{blanchet2018proverif} under the standard Dolev-Yao attacker that fully controls the network (intercept, modify, replay, and inject messages), while cryptographic primitives are modeled as ideal (perfect ML-DSA signatures, perfect ML-KEM encapsulation/decapsulation, collision-resistant hashing, and standard models for HKDF/MAC/AEAD). APQKE handshake ensures secure communication by authenticating the ephemeral ML-KEM key exchange using long-term ML-DSA verification keys and binds the entire handshake context to the hash of the transcript \(TH_1\). The SGN accepts the server after confirming the server's signature on \(TH_1\) and verifying that the server's message is complete. The server accepts the SGN only after verifying the encrypted finished message and then performing decapsulation. The key goal checked in ProVerif is secrecy of the derived \(RK_0\) (and intermediate handshake secrets), where \(RK_0\) is computed from both KEM shared secrets and is transcript-bound as \(RK_0=\mathrm{HKDF}(ss_{SGN_i} \parallel ss_S,\texttt{"dr-root"}\parallel TH_1)\). ProVerif proves secrecy queries (the attacker cannot derive \(RK_0\) or the KEM shared secrets) and authentication correspondence queries (whenever an entity reaches an \emph{accept} event for \(TH_1\), a matching peer \emph{begin} event must have occurred with the same identities and transcript). In the symbolic model, these results rule out classic network attacks against the handshake, such as man-in-the-middle key substitution, replay-based transcript splicing, and impersonation without breaking ML-DSA/ML-KEM or forging the Finished MACs.

DR data channel was modeled in ProVerif as a stateful process initialized from the shared \(RK_0\) and bound to the authenticated handshake by setting \(AD:=TH_1\). Each message key is derived using the symmetric chain KDF and is explicitly treated as one-time: after encryption/decryption, the message key is erased, consistent with the protocol design in which derived \(MK\) values are deleted immediately after use. Each ciphertext is produced with AEAD using associated data that includes the handshake binding and the current header (e.g., \(AAD=(TH_1,\mathrm{hdr})\)), so any header modification or any attempt to transplant a ciphertext into a different handshake transcript is detected by AEAD verification. ProVerif checks (i) secrecy of application plaintexts and per-message keys against an active attacker and (ii) authenticity/integrity via correspondence assertions linking a receiver \emph{decryption-accept} event to a unique sender \emph{encryption-send} event with the same header/counter context, which captures both integrity and replay resistance in the symbolic setting. To model forward secrecy and classical post-compromise recovery, compromise is represented by revealing the current ratchet state at a chosen time. Since past $MK$ values are erased and chains advance, ProVerif can verify that earlier messages remain secret under the symbolic assumptions. In the ProVerif model, even if an attacker knows the current ratchet state, confidentiality of later messages can be restored if a new DH-ratchet step introduces an X25519 shared secret that remains unknown to the attacker. Thus, this is a classical post-compromise recovery result and does not cover a quantum adversary that can break X25519. %Overall, within the ProVerif threat model as described in Section \ref{sec:threat}, DR provides confidentiality, integrity, and channel binding for ongoing traffic, with message-level protection that remains robust against eavesdropping, tampering, and replay attacks on the wire.

% ---------------- Scope-aware compact comparison table ----------------

\begin{table}[tb]
\centering
\scriptsize
\caption{Comparison of APQKE-DR with SOTA smart grid protocols.}
\label{tab:sec-compare}

\setlength{\tabcolsep}{2.5pt}
\renewcommand{\arraystretch}{1.15}

\begin{tabular}{
>{\raggedright\arraybackslash}p{4.85cm}
*{9}{c}
}
\toprule

\textbf{Property} &
\rotatebox{90}{\textbf{APQKE-DR}\textsuperscript{C}} &
\rotatebox{90}{\textbf{TLS 1.3}\textsuperscript{C}} &
\rotatebox{90}{\textbf{Hybrid TLS}\textsuperscript{C}} &
\rotatebox{90}{\textbf{PQXDH}~\cite{kret2023pqxdh}\textsuperscript{H}} &
\rotatebox{90}{\textbf{PQC-SG}~\cite{hafeez2026novel}\textsuperscript{H}} &
\rotatebox{90}{\textbf{Kyber-AMI}~\cite{lagrota2025quantum}\textsuperscript{A}} &
\rotatebox{90}{\textbf{PUF-Auth}~\cite{badar2021identity}\textsuperscript{A}} &
\rotatebox{90}{\textbf{LLRA}~\cite{cheng2024llra}\textsuperscript{A}} &
\rotatebox{90}{\textbf{LSEG}~\cite{zafar2025lseg}\textsuperscript{A}}
\\

\midrule

\multicolumn{10}{l}{\textit{\textbf{Handshake security}}}\\[-0.5mm]

PQ-resistant authenticated key establishment &
$\checkmark$ & $\times$ & $\sim$ & $\sim$ &
$\checkmark$ & $\sim$ & $\times$ & $\times$ & $\times$
\\

Mutual authentication and replay resistance &
$\checkmark$ & $\sim$ & $\sim$ & $\sim$ &
$\sim$ & $\times$ & $\checkmark$ & $\checkmark$ & $\checkmark$
\\

Transcript binding and explicit key confirmation &
$\checkmark$ & $\checkmark$ & $\checkmark$ & $\sim$ &
-- & -- & -- & -- & $\sim$
\\[0.4mm]

\multicolumn{10}{l}{\textit{\textbf{Channel security}}}\\[-0.5mm]

Authenticated application-data channel &
$\checkmark$ & $\checkmark$ & $\checkmark$ & -- &
-- & $\checkmark$ & $\sim$ & $\sim$ & $\checkmark$
\\

Per-message forward secrecy &
$\checkmark$ & $\sim$ & $\sim$ & -- &
-- & $\times$ & $\times$ & $\times$ & $\sim$
\\

Ratcheted messaging &
$\checkmark$ & $\times$ & $\times$ & -- &
-- & $\times$ & $\times$ & $\times$ & $\times$
\\

Classical post-compromise recovery &
$\checkmark$ & $\times$ & $\times$ & -- &
-- & $\times$ & $\times$ & $\times$ & $\times$
\\

Post-quantum post-compromise recovery &
$\times$ & $\times$ & $\times$ & -- &
-- & $\times$ & $\times$ & $\times$ & $\times$
\\[0.4mm]

\multicolumn{10}{l}{\textit{\textbf{Entropy robustness}}}\\[-0.5mm]

Endpoint RNG-failure attacks evaluated &
$\checkmark$ & -- & -- & -- &
-- & -- & -- & -- & --
\\

Entropy/boot mitigation evaluated &
$\checkmark$ & -- & -- & -- &
-- & -- & -- & -- & --
\\[0.4mm]

\multicolumn{10}{l}{\textit{\textbf{Implementation evidence}}}\\[-0.5mm]

Smart-grid endpoint/protocol integration &
$\checkmark$ & -- & -- & -- &
$\sim$ & $\checkmark$ & $\sim$ & $\sim$ & $\checkmark$
\\

Backend/gateway scalability evaluated &
$\times$ & -- & -- & -- &
$\checkmark$ & $\sim$ & -- & -- & --
\\

\bottomrule
\end{tabular}

\vspace{0.5mm}

\parbox{\columnwidth}{\scriptsize
$\checkmark$ = supported; $\sim$ = partial or configuration-dependent; $\times$ = not provided; -- = outside the scheme's stated scope or not evaluated. Superscripts denote primary scope: \textsuperscript{C} complete secure channel, \textsuperscript{H} handshake/session establishment, and \textsuperscript{A} smart grid authentication and key establishment.}
\vspace{-7mm}
\end{table}

Table~\ref{tab:sec-compare} compares handshake security, channel security, entropy robustness, and implementation evidence. APQKE-DR integrates an ML-KEM/ML-DSA handshake with transcript binding and explicit key confirmation, an authenticated data channel with per-message key evolution, classical X25519-based post-compromise recovery, entropy-failure evaluation with tested mitigations, and integration beneath ANSI C12.22 and DNP3. The PQC-SG scheme of~\cite{hafeez2026novel} offers a strong post-quantum smart grid handshake and evaluates graphical processing unit (GPU) based gateway scalability but omits a long-lived ratcheted channel and entropy-failure analysis. TLS variants provide full transport security without continuous ratcheting, while PQXDH establishes an initial secure-messaging secret and leaves channel properties to a later ratchet. Similarly, LSEG~\cite{zafar2025lseg} provides only classical security and lacks post-quantum protection. %Overall, APQKE-DR provides a unified end-to-end design, combining these security and evaluation dimensions into a single experimentally evaluated smart grid secure channel.
%Superscripts indicate whether each work defines a complete channel, a session-establishment handshake, or a smart grid authentication/key-establishment scheme. ``--'' marks properties outside a work's stated scope or evaluation.

%===========================================
% Performance Analysis
%===========================================
\vspace{-2mm}
\subsection{Performance Analysis}
\vspace{-1mm}

\noindent \textbf{APQKE Overhead Analysis: }Table~\ref{tab:p1_bw_lat_one} summarizes the bandwidth and latency overhead of Protocol~1, which completes in three messages (M1 to M3) and outputs the root key $RK_0$ used to seed Algorithm~1. The bandwidth columns report \emph{payload} sizes of the handshake messages produced by our APQKE encoder (excluding transport framing). In this instantiation, the SGN sends M1 and M3 (3{,}690~B and 1{,}144~B), for a total of 4{,}834~B payload, while the server sends M2 (4{,}776~B), yielding 9{,}610~B payload per completed handshake. Message $ M_2$ accounts for 49.7\% of the total payload and includes the server nonce, ML-KEM public key, encapsulated ciphertext, post-quantum authentication, and the Finished MAC. This MAC allows the SGN to verify that SGN and the server derived the same handshake key and agree on the handshake transcript.

Latency is reported as p50/p95 on repeated handshakes with timestamps per-endpoint. On the SGN, we report confirmed completion time as $t_{\mathrm{ack}}-t_{\mathrm{start}}$, where \texttt{ACK} is a benchmark-only confirmation sent by the server after it successfully validates M3 and derives $RK_0$; this avoids undercounting due to asynchronous sends and provides an unambiguous ``channel ready'' completion point. On the server, we report processing time as $t_{\mathrm{done}}-t_{\mathrm{m1\_recv}}$, covering M1 parsing and verification, M2 construction, reception and processing of M3, and final key derivation. In our measurements, the SGN reports 1.990/3.060~ms, while the server reports 0.821/1.113~ms at the p50/p95 levels.

\begin{table}[t]
\centering
\small
\caption{APQKE handshake bandwidth and latency in bytes. %SGN latency is ($t_{\mathrm{ack}}-t_{\mathrm{start}}$); server latency is ($t_{\mathrm{done}}-t_{\mathrm{m1\_recv}}$).
}
\label{tab:p1_bw_lat_one}
\begin{tabular}{p{0.7cm}p{0.7cm}p{0.7cm}p{0.7cm}p{1.2cm}p{2.1cm}}
%\begin{tabular}{|p{1.18cm}|p{0.515cm}|p{1.2cm}|p{0.7cm}|p{1.35cm}|p{1.20cm}|p{0.75cm}|p{0.70cm}|}

\toprule
\textbf{Role (sent)} & \textbf{M1} & \textbf{M2} & \textbf{M3} & \textbf{Total payload} & \textbf{Latency p50/p95 (ms)} \\
\midrule
SGN & 3{,}690 & -- & 1{,}144 & 4{,}834 & 1.990 / 3.060 \\
Server & -- & 4{,}776 & -- & 4{,}776 & 0.821 / 1.113 \\
\midrule
Total & 3{,}690 & 4{,}776 & 1{,}144 & 9{,}610 & -- \\
\bottomrule
\end{tabular}
\vspace{-5mm}
\end{table}

\noindent \textbf{DR Bandwidth Overhead: } We measure the steady-state, per-message bandwidth overhead of the DR protocol at the application/transport boundary. In the DR implementation, each application message is transmitted as two framed records: \texttt{MSG\_HDR} (ratchet header) followed by \texttt{MSG\_DATA} (AEAD ciphertext). To minimize bandwidth in steady state, \texttt{MSG\_HDR} uses the compact header format and carries only a one-byte type tag and a 32-bit message counter $n$, for a 5-byte ratchet-header payload. A longer rekey header (containing $pk_{DH}$ and the previous-chain counter $PN$) is sent only during session bootstrap/DH refresh and is not included in steady-state measurements.

The payload is encrypted with ChaCha20-Poly1305, which expands a plaintext of length $L$ by a 16-byte authentication tag (i.e., ciphertext length is $L+16$). Finally, each record carries a 5-byte framing header (\texttt{type} $\|$ \texttt{len}), so each application message pays 10~bytes of transport framing overhead in total. In the evaluated implementation, the AEAD nonce is not transmitted on the wire; it is derived deterministically from the per-direction message counter (Encode96$(n)$). The on-wire cost per application message is \( W(L) = L + 31 \;\;\text{bytes/message}\). The constant overhead on the wire is \( E(L) = 31 \;\;\text{bytes/message}\), which includes 5 bytes for the ratchet header, 16 bytes for AEAD expansion, and 10 bytes for the transport framing.

Table~\ref{tab:on-wire-bandwidth} reports $W(L)$ and derived overhead/efficiency metrics as a function of payload size. We report steady-state costs and exclude the one-time session bootstrap header observed once per run (an extra \texttt{MSG\_HDR} frame of 46~wire bytes), which is negligible over long sessions.

\begin{table}[tb]
\centering
\small
\caption{DR channel steady-state on-wire overhead and efficiency vs.\ plaintext payload size.}
\setlength{\tabcolsep}{4pt} % optional
\renewcommand{\arraystretch}{1.1} % optional

\begin{tabular}{@{}cccccc@{}}
\toprule
\shortstack{\textbf{Payload}\\\textbf{bytes (L)}} &
\rotatebox{90}{\shortstack{\textbf{Avg wire}\\\textbf{bytes/msg}\\\textbf{(W)}}} &
\rotatebox{90}{\shortstack{\textbf{Overhead}\\\textbf{bytes/msg}\\\textbf{(E)}}} &
\rotatebox{90}{\shortstack{\textbf{Overhead \%}\\\textbf{of Payload (E/L)}}} &
\rotatebox{90}{\shortstack{\textbf{Overhead \%}\\\textbf{of total}\\\textbf{bytes (E/W)}}} &
\rotatebox{90}{\shortstack{\textbf{Efficiency \%}\\\textbf{(L/W)}}} \\
\midrule
16  & 47  & 31 & 193.8 & 66.0 & 34.0 \\
32  & 63  & 31 &  96.9 & 49.2 & 50.8 \\
64  & 95  & 31 &  48.4 & 32.6 & 67.4 \\
128 & 159 & 31 &  24.2 & 19.5 & 80.5 \\
256 & 287 & 31 &  12.1 & 10.8 & 89.2 \\
512 & 543 & 31 &   6.1 &  5.7 & 94.3 \\
\bottomrule
\end{tabular}
\label{tab:on-wire-bandwidth}
\vspace{-6mm}
\end{table}

%-----------------------------------------------------------
%Intergration with ANSI C12.22 and DNP3
%------------------------------------------------------------
\noindent \textbf{Integration with ANSI C12.22 and DNP3: }
To demonstrate deployability within existing smart grid application protocols, we integrate APQKE-DR as a security shim beneath ANSI C12.22~\cite{shees2025performance} and DNP3~\cite{clarke2004practical}. At the shim boundary, each application protocol data unit (PDU) is treated as plaintext: APQKE establishes $(TH_1, RK_0)$, the Double Ratchet is bootstrapped from $RK_0$, and subsequent PDUs are transported as ratcheted AEAD records with associated data $AD := TH_1$. This binds the protected SCADA traffic to the authenticated handshake transcript without altering C12.22/DNP3 format.

Table~\ref{tab:scada-integration} summarizes the costs of ANSI~C12.22 and DNP3. In the current prototype run, the server generates five meter-reading responses consisting of a 64-byte ANSI~C12.19 table payload (Table~27 as given in \cite{shees2025performance}) and a 154-byte DNP3 payload (Group~30 Var~1 analogs and Group~20 Var~1 counters \cite{clarke2004practical}) per reading. The measured processing cost is dominated by the post-quantum handshake: 1.926~ms on the server and 1.872~ms on the client, while DR bootstrap costs 0.136~ms (server) and 0.246~ms (SGN). Protecting the five DNP3 readings (770~B total) costs 0.006~ms for encryption in the sender and 0.004~ms for decryption in the receiver. 

\begin{table}[t]
\centering
\small
\caption{Integrated ANSI~C12.22/DNP3 prototype results with APQKE-DR (five readings).}
\label{tab:scada-integration}
\setlength{\tabcolsep}{7pt}
\renewcommand{\arraystretch}{1.1}
\begin{tabular}{@{}lcc@{}}
\toprule
\textbf{Component (ms)} & \textbf{Server} & \textbf{SGN} \\
\midrule
Handshake & 1.926 & 1.872 \\
DR Bootstrap & 0.136 & 0.246 \\
DR Session Start & 0.102 & 0.101 \\
DR Data Protection (770 B total) & 0.006 & 0.004 \\
\midrule
Total & 2.170 & 2.223 \\
\bottomrule
\end{tabular}
\vspace{-4mm}
\end{table}

%-----------------------------------------------------------
%SOTA Comparison
%------------------------------------------------------------
\noindent \textbf{\textit{SOTA} comparison:}
Table~\ref{tab:pqc_sota} presents a comparison of the client-side execution times of the APQKE handshake with representative PQC-based smart grid protocols. Since reference implementations are publicly available only for \cite{shakib2025amphikey} and \cite{hafeez2026novel}, these two schemes were executed on the same testbed as APQKE to obtain directly comparable measurements. For the remaining protocols, whose implementations are unavailable or measured on different processors, we perform a first-order normalization with respect to CPU clock frequency as \( t_{\mathrm{norm}} = t_{\mathrm{reported}} \times \frac{f_{\mathrm{reported}}}{2.4~\mathrm{GHz}}\)\footnote{This normalization is approximate and does not account for differences in instruction set architecture, microarchitecture, memory hierarchies, or implementation-specific optimizations.}.

On the 2.4~GHz Cortex-A76 platform, the APQKE handshake attains a median SGN completion time of $1.99$~ms. Its normalized client-side latency is lower than that of the three schemes of \cite{shekhawat2024quantum1}, \cite{shekhawat2024quantum2} ($2.04$ and $2.39$~ms), \cite{hafeez2026novel} ($2.29$~ms), and the authenticated configuration of \cite{shakib2025amphikey} ($4.52$~ms). The scheme \cite{bera2024securing} reports a lower normalized execution time of $1.17$~ms, but it omits several critical security properties. Overall, APQKE resides within the performance envelope of contemporary PQC-based smart grid handshakes. The comparison in Table~\ref{tab:pqc_sota} is restricted to timing, as the schemes differ in parameterization and in the extent of their security functionality. %In particular, the complete APQKE-DR construction additionally offers a transcript-bound ratcheted channel and includes an empirical evaluation under endpoint entropy-failure scenarios.

\begin{table}[tb]
\centering
\caption{Approximate comparison of client-side timing using
first-order frequency normalization to 2.4~GHz.}
\vspace{-2mm}
\label{tab:pqc_sota}
\small
\begin{tabular}{llrr}
\hline
Protocol & Platform & $f_{\mathrm{rep.}}$ (GHz) &
$t_{\mathrm{norm}}$ (ms) \\
\hline
~\cite{bera2024securing}
    & RPi 4 (A72) & 1.40 & 1.17 \\
~\cite{shekhawat2024quantum1}
    & RPi 4 (A72) & 1.40 & 2.04 \\
~\cite{shekhawat2024quantum2}
    & RPi 4 (A72) & 1.40 & 2.39 \\
~\cite{hafeez2026novel}
    & RPi CM5 (A76) & 2.40 & 2.29 \\
~\cite{shakib2025amphikey}
    & RPi CM5 (A76) & 2.40 & 4.52 \\
APQKE
    & RPi CM5 (A76) & 2.40 & 1.99 \\
\hline
\end{tabular}
\vspace{-6mm}
\end{table}

%-----------------------------------------------------------
%TPM vs Non-TPM
%------------------------------------------------------------
\vspace{-2mm}
\subsection{Hardware-Anchored Entropy on the SGN (TPM 9670)}
\label{sec:tpm_entropy}
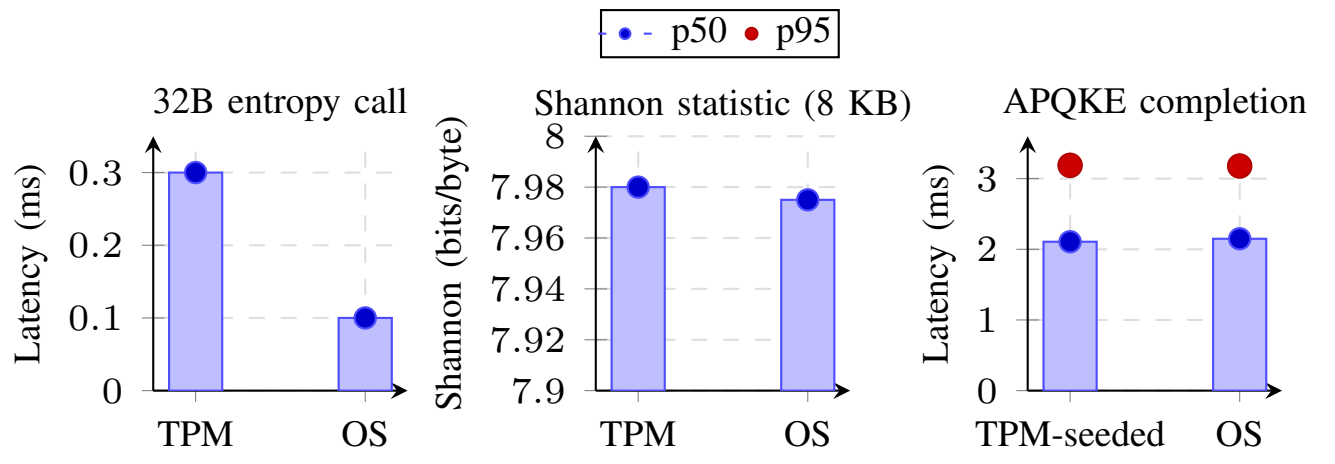
\begin{figure}[t]
\centering

% ---- measured values (NO digits in macro names!) ----
\newcommand{\TPMentropyMs}{0.3}
\newcommand{\OSentropyMs}{0.1}

\newcommand{\TPMentropyBits}{7.98}
\newcommand{\OSentropyBits}{7.975}

\newcommand{\PoneN}{2000}

% Protocol-1 SGN confirmed completion from your CSV logs (ms)
\newcommand{\TPMpOnePFifty}{2.107}
\newcommand{\TPMpOnePNinetyFive}{3.188}
\newcommand{\OSpOnePFifty}{2.149}
\newcommand{\OSpOnePNinetyFive}{3.178}

\begin{tikzpicture}

\begin{groupplot}[
  group style={group size=3 by 1, horizontal sep=1.35cm},
  width=0.18\textwidth, height=0.18\textwidth,
  axis lines=left,
  grid=major,
  major grid style={dashed, gray!25},
          ylabel style={yshift=-3pt},
        xlabel style={yshift=6pt},
  tick label style={font=\scriptsize},
  label style={font=\scriptsize},
  title style={font=\scriptsize, yshift=-1.5ex},
  enlarge x limits=0.25,
  ymajorgrids=true,
  xtick=data,
]

% -------------------- Plot 1: 32B entropy call latency --------------------
\nextgroupplot[
  title={32B entropy call},
  ylabel={Latency (ms)},
  symbolic x coords={TPM, OS},
  xticklabels={TPM,OS},
  ymin=0, ymax=0.35,
  ytick={0,0.1,0.2,0.3},
]

\addplot+[ybar, bar width=10pt, fill=blue!25, draw=blue!70] coordinates {
  (TPM,\TPMentropyMs)
  (OS,\OSentropyMs)
};

% -------------------- Plot 2: Shannon entropy estimate --------------------
\nextgroupplot[
  title={Shannon statistic (8 KB)}, xshift=-3pt,
  ylabel={Shannon (bits/byte)},
  symbolic x coords={TPM,OS},
  xticklabels={TPM,OS},
  ymin=7.90, ymax=8.00,
  ytick={7.90,7.92,7.94,7.96,7.98,8.00},
]

\addplot+[ybar, bar width=10pt, fill=blue!25, draw=blue!70] coordinates {
  (TPM,\TPMentropyBits)
  (OS,\OSentropyBits)
};

% -------------------- Plot 3: Protocol 1 completion (SGN) --------------------
\nextgroupplot[
  title={APQKE completion}, xshift=-5pt,
  ylabel={Latency (ms)},
  symbolic x coords={TPM,OS},
  xticklabels={TPM-seeded,OS},
  ymin=0, ymax=3.6,
  ytick={0,1,2,3},
        legend style={font=\scriptsize, at={(-1.21, 1.5)}, anchor=north,
                      legend columns=2, inner sep=-0.7pt, row sep=-2pt,
                      column sep=1.5pt, nodes={scale=1.0, transform shape},
                      legend image post style={scale=0.5}},
]

% p50 as bars
\addplot+[ybar, bar width=10pt, fill=blue!25, draw=blue!70] coordinates {
  (TPM,\TPMpOnePFifty)
  (OS,\OSpOnePFifty)
};
\addlegendentry{p50}

% p95 as red markers
\addplot+[only marks, mark=*, mark size=2.2pt, draw=red!70!black, fill=red!70!black] coordinates {
  (TPM,\TPMpOnePNinetyFive)
  (OS,\OSpOnePNinetyFive)
};
\addlegendentry{p95}

\end{groupplot}
\end{tikzpicture}
\vspace{-7mm}
\caption{Entropy-source comparison on the Raspberry Pi SGN (SPI TPM 2.0 vs OS CSPRNG): 32-byte call latency and 8 KB output-distribution sanity check; APQKE p50/p95 from \PoneN\ confirmed handshakes.}
\label{fig:tpm_entropy_char}
\vspace{-6mm}
\end{figure}

\vspace{-2mm}
Since endpoint entropy failures can undermine key establishment on embedded devices, we evaluated a TPM~2.0 module (TPM~9670 over SPI \cite{festa2025anchoring}) on the SGN as an additional hardware-based entropy source. The TPM is \emph{not} used as a bulk randomness generator; instead, it is queried only to obtain a small amount of seed data at the start of the session. Fig.~\ref{fig:tpm_entropy_char} compares \texttt{TPM2\_GetRandom} output with the OS CSPRNG on the Raspberry Pi SGN. A 32\,B TPM entropy request takes 0.3 \,ms, while an OS RNG read completes in 0.1 \,ms. As a lightweight sanity check, we collect 8\,KB of output from each source and compute Shannon entropy, obtaining 7.980\,bits/B (TPM) and 7.975\,bits/B (OS), close to 8.0\,bits/B maximum.

To quantify whether TPM seeding impacts APQKE latency, we run the \emph{APQKE protocol with TPM-seeded entropy} and with OS-only entropy for $2000$ confirmed handshakes (N) per setting. We report the SGN confirmed completion time as $(t_{\mathrm{ack}}-t_{\mathrm{start}})$, where the server emits \texttt{ACK} after successfully processing $M_3$ and deriving $RK_0$. The SGN p50/p95 completion times remain comparable (TPM-seeded: 2.107/3.188\,ms; OS-only: 2.149/3.178\,ms), indicating that injecting a small hardware-anchored seed can be done without a large per-handshake latency penalty in our prototype.

% ---------- Figure ----------

%-----------------------------------------------------------
%Intergration with ANSI C12.22 and DNP3
%------------------------------------------------------------
\vspace{-4mm}
\subsection{EAS Analysis}\label{subsec:EAS_results}
\vspace{-2mm}

\noindent \textbf{Early-boot entropy hole (BWSR):} We have evaluated the early-boot entropy hole by starting APQKE at a controlled delay $d$ after reset under the injected weak-RNG model described in Section~\ref{sec:eas-boot}, and measuring the attacker's key-recovery success probability $p_{\mathrm{KR}}(d)$. For each delay setting, we record $R$ independent Protocol~1 transcripts and run the offline seed-recovery procedure. We then compute \(p_{\mathrm{KR}}(d)= \frac{\text{sessions in which the attacker recovers }RK_0}{R}\). To characterize the observed transition, we define \(T_{0.5}\) as the earliest tested delay $d$ at which $p_{\mathrm{KR}}(d)$ falls below the fixed threshold $\tau=0.5$. Intuitively, $T_{0.5}$ separates the tested region in which recovery is more likely than not from the region in which it is less likely than not. This metric characterizes the controlled injected model; it is not an estimate of the time required for the underlying operating-system CSPRNG to become initialized.

To characterize the feasibility of the offline phase, we also report the implied seed-search work factor\footnote{The attacker tests possible early-boot seeds $s$ from a space of size $2^b$. Each attempt requires $t_{\mathrm{try}}$ seconds. The work factor is the expected number of trials or total time required to identify a consistent seed.}: for an effective early-boot seed space $2^b$ and measured per-try simulation time $t_{\mathrm{try}}$, the expected search time is \(\mathbb{E}[T_{\mathrm{search}}]\approx 2^{b-1}t_{\mathrm{try}}\).

On the SGN, we start APQKE after a fixed delay $d$ and set the weak-RNG interval to $T_{\mathrm{inj}}$. For each $d$, we collect $R=20$ independent handshakes and run the offline attacker on each transcript. We evaluate: (i) \emph{OS-only}, where we inject the early-boot RNG failure at the SGN RNG interface, and (ii) \emph{TPM-seeded}, where hardware-derived seed material is mixed into the per-session RNG state before APQKE.

\begin{figure*}[htb]
\centering
\begin{tikzpicture}

% ---------- Panel sizing (fit 5 in one row) ----------
\def\wbig{0.27\linewidth}
\def\wsm {0.155\linewidth}   % was 0.17
\def\wmed{0.155\linewidth}   % was 0.17
\def\hpanel{4.0cm}
\def\xsep{1pt}               % was -4pt

% =========================================================
% (a) p_KR(d) vs delay
% =========================================================
\begin{axis}[
  name=panA,
  width=\wbig, height=\hpanel,
  title={(a) Key-recovery probability},
  title style={font=\scriptsize},
  xlabel={Handshake start delay after boot (ms)},
  ylabel={$p_{\mathrm{KR}}(d)$},
  xmin=0, xmax=3000,
  ymin=0, ymax=1.25,
  xtick={0,500,1000,1500,2000,2500,3000},
  ytick={0.25,0.5,0.75,1.0,1.25},
  yticklabels={0.25,0.5,0.75,1,1.25},
  grid=both,
  major grid style={black!15},
  ylabel style={yshift=-3pt},
  minor grid style={black!7},
  tick label style={font=\scriptsize},
  label style={font=\scriptsize},
  xticklabel style={rotate=45, anchor=east},
  legend style={
    font=\scriptsize,
    draw=black!20,
    fill=white,
    fill opacity=0.9,
    text opacity=1,
    at={(0.02,0.98)},
    anchor=north west,
    row sep=1pt
  },
  legend cell align=left,
]

\addplot[draw=none, fill=black!8, forget plot] coordinates
  {(0,0) (2000,0) (2000,1.25) (0,1.25)} \closedcycle;

\addplot[thick, densely dotted, forget plot] coordinates {(0,0.5) (3000,0.5)};

\addplot[very thick, mark=*, mark size=1.2pt] coordinates {
  (0,1) (500,1) (1000,1) (1500,1)
  (2000,0) (2500,0) (3000,0)
};

\addplot[very thick, dashed, mark=square*, mark size=1.2pt] coordinates {
  (0,0) (500,0) (1000,0) (1500,0)
  (2000,0) (2500,0) (3000,0)
};

\addplot[thick, dashdotted, forget plot] coordinates {(2000,0) (2000,1.25)};

\draw[<->, >=Stealth, thick] (axis cs:0,1.20) -- (axis cs:2000,1.20);
\node[font=\scriptsize, fill=none, inner sep=1pt]
  at (axis cs:1000,1.08) {Weak-RNG interval};

\end{axis}

% =========================================================
% (b) Seed uncertainty vs observed prefix length
% =========================================================
\begin{axis}[
  name=panB,
  at={(panA.outer north east)}, anchor=outer north west,
  xshift=\xsep,
  width=\wbig, height=\hpanel,
  title={(b) Prefix leakage},
  title style={font=\scriptsize},
  xlabel={Observed nonce prefix length (bytes)},
  ylabel={Seed uncertainty (bits)},
  xmin=0, xmax=16,
  ymin=7.5, ymax=16.5,
  xtick={0,2,4,6,8,10,12,14,16},
  ytick={8,9,10,11,12,13,14,15,16},
  ylabel style={yshift=-3pt},
  grid=both,
  major grid style={black!15},
  minor grid style={black!7},
  tick label style={font=\scriptsize},
  label style={font=\scriptsize},
]
\addplot+[const plot, very thick] coordinates {
  (0,16) (1,16) (1,8) (16,8)
};
\end{axis}

% =========================================================
% (c) Structural key recovery (bar)
% =========================================================
\begin{axis}[
  name=panC,
  at={(panB.outer north east)}, anchor=outer north west,
  xshift=\xsep,
  width=\wsm, height=\hpanel,
  title={(c) Structural key recovery},
  title style={font=\scriptsize},
  ybar,
  bar width=8pt,
  ymin=0, ymax=1.5,
  ylabel={$p_{\mathrm{KR}}$},
  symbolic x coords={OS,TPM,TPM-PCR},
  xtick=data,
  xticklabels={OS,TPM,TPM-PCR},
  xticklabel style={font=\scriptsize, rotate=45, anchor=east},
  ytick={0,0.5,1.0,1.5},
  grid=both,
  major grid style={black!15},
  minor grid style={black!7},
  ylabel style={yshift=-3pt},
  tick label style={font=\scriptsize},
  label style={font=\scriptsize},
  enlarge x limits=0.35,
]
\addplot coordinates {
  (OS,1.0)
  (TPM,0.95)
  (TPM-PCR,0)
};

\node[
  rotate=90,
  anchor=west,
  font=\scriptsize\bfseries,
  inner sep=1pt
] at (axis cs:TPM-PCR,0.06) {Blocked};

\end{axis}

% =========================================================
% (d) Output collision check (bar)
% =========================================================
\begin{axis}[
  name=panD,
  at={(panC.outer north east)}, anchor=outer north west,
  xshift=\xsep,
  width=\wsm, height=\hpanel,
  title={(d) Output collision check},
  title style={font=\scriptsize},
  ybar,
  bar width=8pt,
  ymin=0, ymax=2200,
  ylabel={Unique $RK_0[0..7]$ prefixes},
  symbolic x coords={TPM,OS},
  xtick=data,
  xticklabels={TPM,OS},
  ylabel style={yshift=-3pt},
  xticklabel style={font=\scriptsize, rotate=45, anchor=east},
  ytick={0,500,1000,1500,2000},
  grid=both,
  major grid style={black!15},
  minor grid style={black!7},
  tick label style={font=\scriptsize},
  label style={font=\scriptsize},
  enlarge x limits=0.35,
  %nodes near coords,
  every node near coord/.append style={font=\tiny},
]
\addplot coordinates {(TPM,2000) (OS,2000)};
\end{axis}

% =========================================================
% (e) Hamming-weight boxplot (same row)
% =========================================================
\begin{axis}[
  name=panE,
  at={(panD.outer north east)}, anchor=outer north west,
  xshift=\xsep,
  width=\wmed, height=\hpanel,
  title={(e) HW sanity check},
  ylabel style={yshift=-3pt},
  title style={font=\scriptsize},
  boxplot/draw direction=y,
  ylabel={\shortstack{Hamming weight of\\first 8 bytes of $RK_0$}},
  xtick={1,2},
  xticklabels={OS,TPM},
  xticklabel style={font=\scriptsize, rotate=45, anchor=east},
  ymin=15, ymax=50,
  grid=both,
  major grid style={black!15},
  minor grid style={black!7},
  tick label style={font=\scriptsize},
  label style={font=\scriptsize},
]

\addplot+[
  boxplot prepared={
    lower whisker=20,
    lower quartile=29,
    median=32,
    upper quartile=35,
    upper whisker=46
  },
  very thick,
] coordinates {};

\addplot+[
  boxplot prepared={
    lower whisker=19,
    lower quartile=29,
    median=32,
    upper quartile=35,
    upper whisker=46
  },
  very thick,
] coordinates {};

\end{axis}

\end{tikzpicture}
\vspace{-8mm}
\caption{(a) $p_{\mathrm{KR}}(d)$ vs handshake start delay, (b) low-byte LCG aliasing: one nonce byte reduces $2^{16}$ states to a $2^8$-state equivalence class, (c) structural key recovery under RNG-path subversion, (d) output collision check: all $R=2000$ observed $RK_0[0..7]$ prefixes were distinct in both configurations, and (e) HW sanity check for $RK_0[0..7]$ over $R=2000$ handshakes.}
\label{fig:four-panels}
\vspace{-5mm}
\end{figure*}
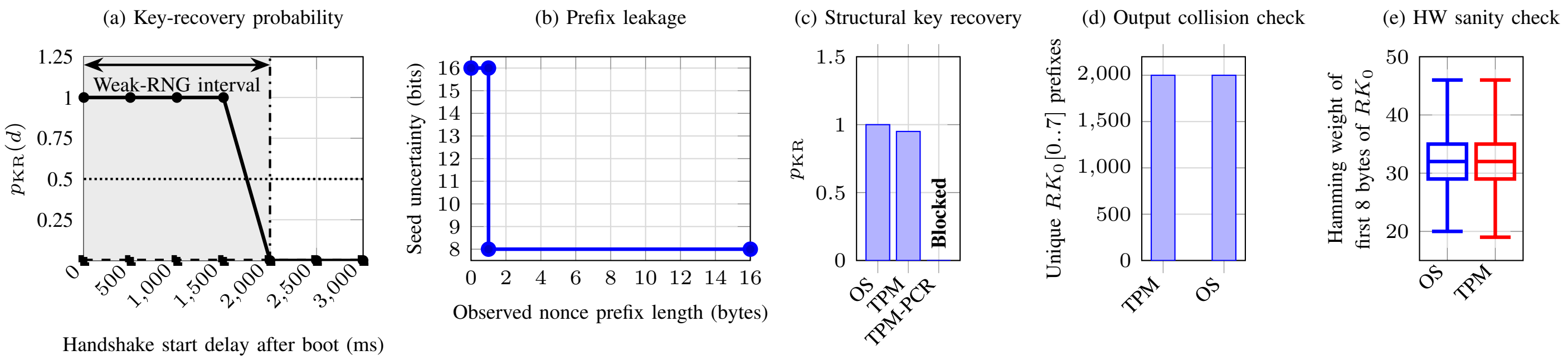

Fig.~\ref{fig:four-panels}(a) plots $p_{\mathrm{KR}}(d)$ versus handshake start delay. In the OS-only configuration, key recovery remains reliable for handshakes starting within the weak-RNG interval and drops below $\tau=0.5$ at about $d=2000~\mathrm{ms}$. At this point, recovery occurs in only a few of the $R=20$ trials, and no recovery is seen at larger delays. Under this threshold, the injected experiment yields \(T_{0.5}^{\mathrm{inj}}\approx2000~\mathrm{ms}\). In the TPM-seeded configuration, the attacker never recovers a root key ($p_{\mathrm{KR}}(d)=0/20$ for all $d$), so \(T_{0.5}^{\mathrm{tpm}}=0~\mathrm{ms}\).

This controlled experiment shows that a weak early-boot RNG state, for example, from repeated resets or incomplete entropy initialization, can let a recorded APQKE transcript support offline reconstruction of $RK_0$. The $\approx2000~\mathrm{ms}$ transition is specific to our injected failure setup and is not a measurement of a Raspberry Pi OS RNG vulnerability or initialization time. TPM seeding removes the predictable boot-time dependence in this model.

%The results were obtained on a Linux-based embedded platform, where early-boot entropy is affected by kernel data, interrupts, and peripheral activity. To simulate worst-case conditions, the system is rebooted before each test, and APQKE is initialized early in the boot to minimize interference. A more controlled evaluation would involve running the protocol on bare metal to better analyze the entropy-seeding interface and assess attacker replay capabilities under specific constraints.

\noindent \textbf{Results: RNG path subversion (LCG):} We have evaluated RNG-path subversion by interposing a malicious shim (i.e., an interposition layer that replaces RNG API outputs) at the SGN-side RNG API and forcing all device-side randomness consumed by APQKE protocol to be generated by a deterministic 16-bit LCG. We have executed $R=2000$ APQKE handshakes per SGN configuration and logged the public transcript (in particular, $M_1$ containing $N_{SGN_i}$) and the resulting root-key material. We compared three builds on the same platform: (i) \emph{OS-only} randomness (\texttt{USE\_TPM=0}) and (ii) \emph{TPM-seeded} randomness (\texttt{USE\_TPM=1}), and (iii) \emph{TPM-PCR} (\texttt{TPM-PCR} in Fig.~\ref{fig:four-panels}(c)), which adds a \emph{fail-closed} TPM gate before APQKE use any randomness. Concretely, the SGN requests a short secret $K_{\text{gate}}$ (16 bytes) from the TPM; this value is \emph{sealed} to the platform’s measured-boot state (TPM-PCRs are registers that store hashes of boot components). If the boot state matches a known-good baseline, the TPM releases $K_{\text{gate}}$; otherwise, the request fails. The SGN then XOR-mixes $K_{\text{gate}}$ into the per-handshake seed used by APQKE (nonce generation, ML-KEM key generation, and encapsulation coins) and aborts the handshake if $K_{\text{gate}}$ cannot be obtained. In (i) and (ii), the LCG shim is positioned \emph{after} the seeding source (OS or TPM), so it replaces the bytes returned to APQKE regardless of where the seed was originally sourced.

The shim maintains a nominal 16-bit internal register as in Fig.~\ref{fig:four-panels}(b) and updates it using $\mathrm{st}_{j+1} =(a\cdot \mathrm{st}_j+c)\bmod 2^{16}$, with $(a,c)=(25173,13849)$, outputting each byte as the low 8 bits of the updated state. Because the protocol observes only byte outputs, the low-byte projection produces only $2^8$ distinct observable streams. After observing the first byte of $N_{SGN_i}$, the attacker directly identifies the corresponding observable stream. Although $2^8$ internal states remain consistent with that byte, they form an equivalence class whose elements generate an identical output stream and therefore provide no additional attack resistance. Fig.~\ref{fig:four-panels}(b) visualizes this internal-state aliasing. This LCG thus serves as an intentionally weak, 8-bit-observable diagnostic model for validating end-to-end key-recovery propagation and for evaluating the effectiveness and placement of mitigation mechanisms.

Given a recorded transcript, the attacker identifies an observable stream representative consistent with the observed $N_{SGN_i}$ prefix and replays the APQKE protocol offline while forcing the same deterministic RNG stream. This regenerates the SGN's RNG dependent computations for nonce generation, ephemeral ML-KEM key generation, and encapsulation randomness, enabling the reconstruction of both ML-KEM shared secrets and, thus, the APQKE-derived root key. The attacker does not need to recover the unique 16-bit internal state because every state within the identified equivalence class generates the same protocol visible byte stream.

Across the $R=2000$ logged runs, the attacker succeeds with overwhelming probability, as shown in Fig.~\ref{fig:four-panels}(c), for the OS-only and TPM-seeded builds. Importantly, TPM seeding does not reduce $p_{\mathrm{KR}}$ in this threat model because subversion occurs \emph{after} the TPM/OS entropy source; once the RNG API is replaced, protocol consumes attacker-controlled bytes independent of upstream entropy quality. In contrast, under TPM PCRs gating, LCG-shim firmware changes the boot measurements reflected in TPM PCRs, so the TPM refuses to release $K_{\mathrm{gate}}$. If $K_{\mathrm{gate}}$ is unavailable, the TPM PCR build fails closed, the SGN aborts before finishing APQKE, and no \emph{accepted} sessions are established under the shim image.

A na\"ive output-only check is to count repeated $RK_0$ values across handshakes. In our logs ($R=2000$ per configuration), all observed $RK_0[0..7]$ prefixes were distinct for both OS-only and TPM-seeded runs, as shown in Fig.~\ref{fig:four-panels}(d). Server-generated values and public transcript vary across sessions, so HKDF can produce distinct, apparently well-distributed root keys even when the SGN-side RNG is restricted to only $2^8$ observable streams. For any recorded transcript, however, the attacker still enumerates at most these $2^8$ possible streams of bytes and verifies the correct one using the public nonce and KEM ciphertexts. Therefore, absence of repeated $RK_0$ prefixes does not imply resistance to transcript-guided root-key recovery.%
\footnote{TPM-PCR blocks the modified image before key establishment; therefore, no attacked sessions yield $RK_0$ samples, and the corresponding metrics are reported as blocked.}

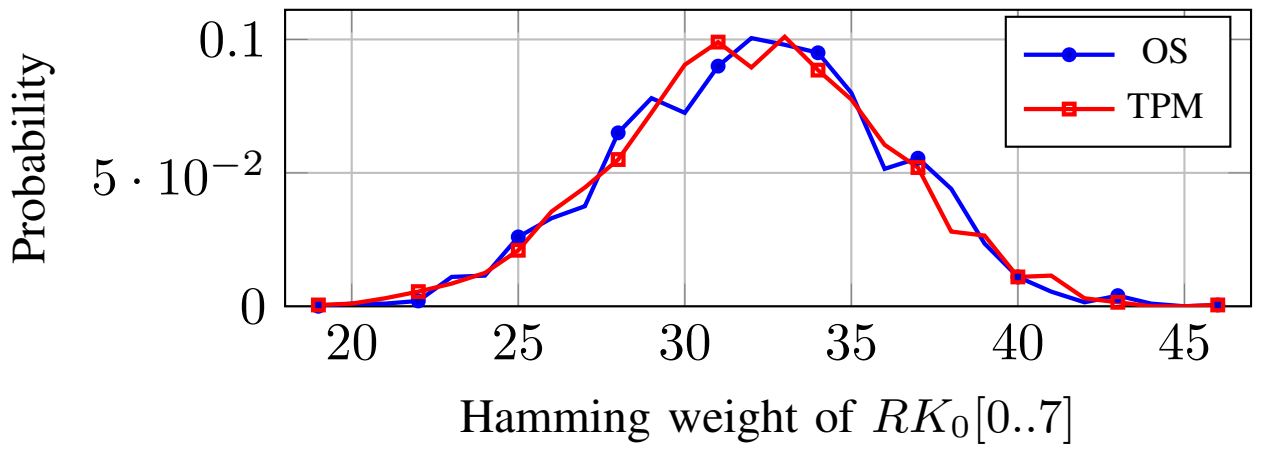
\begin{figure}[t]
\centering
\begin{tikzpicture}
\begin{axis}[
    width=0.9\linewidth,
    height=0.4\linewidth,
    xlabel={\small Hamming weight of $RK_0[0..7]$},
    ylabel={\small Probability},
    xmin=18, xmax=47,
    ymin=0,
    grid=both,
    legend style={at={(0.745,0.98)}, anchor=north west},
]
% OS-only empirical PMF (R=2000)
\addplot+[thick, mark=*, mark size=1.0pt, mark repeat=3] coordinates {
(19,0.0000) (20,0.0010) (21,0.0010) (22,0.0020) (23,0.0110) (24,0.0115)
(25,0.0260) (26,0.0330) (27,0.0375) (28,0.0650) (29,0.0780) (30,0.0725)
(31,0.0900) (32,0.1005) (33,0.0980) (34,0.0950) (35,0.0800) (36,0.0515)
(37,0.0555) (38,0.0440) (39,0.0235) (40,0.0110) (41,0.0055) (42,0.0015)
(43,0.0040) (44,0.0010) (45,0.0000) (46,0.0005)
};
\addlegendentry{\scriptsize OS}

% TPM-seeded empirical PMF (R=2000)
\addplot+[thick, mark=square, mark size=1.0pt, mark repeat=3] coordinates {

(19,0.0005) (20,0.0010) (21,0.0030) (22,0.0055) (23,0.0085) (24,0.0125)
(25,0.0210) (26,0.0355) (27,0.0445) (28,0.0550) (29,0.0725) (30,0.0905)
(31,0.0990) (32,0.0895) (33,0.1010) (34,0.0885) (35,0.0775) (36,0.0605)
(37,0.0520) (38,0.0280) (39,0.0265) (40,0.0110) (41,0.0115) (42,0.0030)
(43,0.0015) (44,0.0000) (45,0.0000) (46,0.0005)
};
\addlegendentry{\scriptsize TPM}
\end{axis}
\end{tikzpicture}
\vspace{-3mm}
\caption{OS and TPM empirical probability mass function of the Hamming weight of $RK_0[0..7]$ over $R=2000$ handshakes.}
\label{fig:rk0-hw-pmf}
\vspace{-6mm}
\end{figure}

To illustrate how distributional sanity checks can be misleading, Fig.~\ref{fig:four-panels}(e) and Fig.~\ref{fig:rk0-hw-pmf} show that the Hamming weight of $RK_0[0..7]$ looks statistically similar between OS-only and TPM-seeded runs. Despite this apparent normality, deterministic replay still reconstructs $RK_0$ offline because the weakness is \emph{structural}, not a detectable bias in $RK_0$ outputs. However, the TPM-PCR mitigation does not rely on output ``sanity''; it fails closed by refusing to establish any session under shim image.

Recovery of $RK_0$ permits forgery of the $K_{\mathrm{boot}}$ authenticators and therefore enables active interference with DR bootstrap. It does not by itself reveal chain keys from an already completed honest bootstrap, because initialization also mixes the fresh X25519 secret $Z_0$. If the RNG subversion affects X25519 key generation, DH secrets become predictable and post-compromise recovery is compromised.

\vspace{-2mm}
\subsection{Discussion and Limitations}
\vspace{-1mm}
APQKE-DR offers a practical post-quantum-initialized secure channel for smart grid endpoints. ProVerif analysis shows the APQKE handshake keeps $RK_0$ secret and resists active attacks like key substitution and replay. Since APQKE runs only during setup (e.g., initial association or after state loss), its communication overhead is spread over long-term operation. In steady state, the Double Ratchet adds 31~B/message overhead, notable only for very small PDUs and negligible as payloads grow. ANSI~C12.22 and DNP3 shim tests show the post-quantum handshake dominates setup cost, while ratcheted protection stays lightweight and requires no app. layer changes.

Although we implement and evaluate APQKE-DR in a smart grid setting, the core construction is not tied to ANSI~C12.22, DNP3, or any specific smart grid architecture. It applies to deployments with provisioned endpoint identities, managed backends, long-lived encrypted sessions, potential reboot or state loss, and a need for post-quantum initial key establishment combined with lightweight per-message key evolution. Examples include industrial control and SCADA, water and gas metering, remote environmental monitoring, edge-IoT and medical devices, transportation infrastructure, and other critical infrastructure. In these settings, APQKE-DR can protect existing application payloads using a security layer, as in our smart grid prototypes. However, domain-specific validation is required for each domain before large-scale implementation.

EAS results show that long-session security depends on the quality of endpoint randomness and RNG-path integrity. In the controlled early-boot model, $RK_0$ is recoverable during the injected weak-RNG window, with $T_{0.5}^{\mathrm{inj}}\approx 2000$~ms at $\tau=0.5$. TPM seeding blocks recovery across tested delays with minimal latency impact, but cannot stop downstream RNG substitution, where APQKE accepts attacker-controlled bytes and output-only checks can still look normal.

%===========================================
% Conclusions
%===========================================
\vspace{-2mm}
\section{Conclusions}\label{sec:con}
\vspace{-1mm}

We presented APQKE-DR and evaluated its performance and resilience to endpoint randomness failures using EAS as proposed. The protocol combines authenticated post-quantum key establishment with an X25519 Double Ratchet, whose post-compromise recovery relies on classical assumptions. The prototype integrates beneath ANSI~C12.22 and DNP3 without changing application payload formats. It achieves a median confirmed handshake completion time of 1.990~ms and adds 31~B per message. Under fresh randomness and ideal cryptographic primitives, ProVerif verifies the modeled secrecy and authentication properties.

The EAS experiments show that these properties can be undermined when endpoint randomness fails. In the tested cases, predictable randomness allowed recovery of the initial root key from public handshake transcripts. TPM seeding prevented recovery in the tested early-boot cases but did not prevent RNG substitution after seeding. Measured-boot gating blocked the tested modified firmware from establishing a session.

\vspace{-3mm}
\vspace{-2mm}
%%%%%%%%%%%%%%%%%%%%%%%%%%%%%%%%%%%%%%%%%%%%%%%%%%%%%%%%%%%%%%%%%%%%%%%%%%%%%%%%%%%%%%%%%%%%%%%%%%%%%

\bibliographystyle{IEEEtran}
\bibliography{references}  % your file name, without the .bib extension

% that's all folks
\end{document}